\documentclass[aps, twocolumn, superscriptaddress, amsmath, amssymb, floatfix, prb, longbibliography, nofootinbib, notitlepage]{revtex4-2}
\usepackage{amsbsy}
\usepackage{amsfonts}
\usepackage{amsmath}
\usepackage{amstext}
\usepackage{amsthm}
\usepackage{amssymb}
\usepackage{graphicx}
\usepackage{epsfig}
\usepackage{epstopdf}
\usepackage{enumerate}
\usepackage{bm}
\usepackage{bbm}
\usepackage{color}
\definecolor{prlblue}{rgb}{0.18,0.19,0.57}
\usepackage{multirow}
\usepackage[colorlinks]{hyperref}
\usepackage{cleveref}
\usepackage{ifthen}
\usepackage{tikz}
\usepackage{multirow}
\usepackage{braket}
\usepackage{comment}
\usetikzlibrary{patterns.meta} 
\usetikzlibrary{calc}

\newcommand{\1}{\text{\uppercase\expandafter{\romannumeral1}}}
\newcommand{\2}{\text{\uppercase\expandafter{\romannumeral2}}}
\newcommand{\3}{\text{\uppercase\expandafter{\romannumeral3}}}
\newcommand{\4}{\text{\uppercase\expandafter{\romannumeral4}}}
\newcommand{\5}{\text{\uppercase\expandafter{\romannumeral5}}}
\newcommand{\6}{\text{\uppercase\expandafter{\romannumeral6}}}
\makeatletter

\newsavebox{\@brx}
\newcommand{\llangle}[1][]{\savebox{\@brx}{\(\m@th{#1\langle}\)}%
  \mathopen{\copy\@brx\kern-0.5\wd\@brx\usebox{\@brx}}}
\newcommand{\rrangle}[1][]{\savebox{\@brx}{\(\m@th{#1\rangle}\)}%
  \mathclose{\copy\@brx\kern-0.5\wd\@brx\usebox{\@brx}}}
\makeatother

\def\k{\bm{k}}

\renewcommand{\L}{\mathcal{L}}

\newcommand{\be}{\begin{equation}}
\newcommand{\ee}{\end{equation}}

\begin{document}
\graphicspath{{figures/}}

\title{Single-band square lattice Hubbard model from twisted bilayer C$_{568}$}

\author{Toshikaze Kariyado}
\affiliation{International Center for Materials Nanoarchitectonics (WPI-MANA),
National Institute for Materials Science, Tsukuba 305-0044, Japan}

\author{Ashvin Vishwanath}
\affiliation{Department of Physics, Harvard University, Cambridge, MA 02138, USA}

\author{Zhu-Xi Luo}
\affiliation{Department of Physics, Harvard University, Cambridge, MA 02138, USA}
\affiliation{School of Physics, Georgia Institute of Technology, Atlanta, Georgia 30332, USA}

\date{\today}

\begin{abstract}
We propose twisted homobilayer of a carbon allotrope, C$_{568}$, to be a promising platform to realize controllable square lattice single-band extended Hubbard model. This setup has the advantage of a widely tunable $t'/t$ ratio without adding external fields, and the intermediate temperature $t\ll T\ll U$ regime can be easily achieved. We first analyze the continuum model obtained from symmetry analysis and first-principle calculations, and calculate the band structures. Subsequently, we derive the corresponding tight-binding models and fit the hopping parameters as well as the Coulomb interactions. When displacement field is applied, anisotropic nearest neighbor hoppings can further be achieved. 
If successfully fabricated, the device could be an important stepping stone towards understanding 
high-temperature superconductivity. 
\end{abstract}

\maketitle

\setcounter{tocdepth}{1} 
%{\small \tableofcontents }
{
  \hypersetup{linkcolor=magenta}
  \tableofcontents
}

\section{Introduction}
%One of the biggest dreams 

One of the most highly prized goals in condensed matter physics is to realize controllable high-temperature superconductivity. The cuprate materials comprise a famous family of high-temperature superconductors \cite{keimer2015quantum}, which are fascinating but also challenging to understand. Recently, the rise of van der Waals heterostructures as highly controllable quantum simulators of various quantum systems \cite{kennes2021moire}, especially Hubbard models \cite{PhysRevLett.121.026402,tang2020simulation} naturally lead us to wonder: can van der Waals materials simulate cuprate physics and particularly high-temperature superconductivity? 

It is widely believed that the simplest effective model that captures essential cuprate physics is that of the single-band square-lattice Hubbard model , see for example \cite{review_cuprates}. To simulate the latter, a Moiré superlattice with square unit cells is necessary. This can be achieved through various setups, including twisted homobilayers of square lattice materials, heterobilayers of square lattice materials with mismatched lattice constants, two-dimensional materials from other crystal groups twisted at special angles, or multilayer heterostructures. For example, recently the possibility of using $90^{\circ}$-twisted homobilayer of the GeSe family to simulate cuprate physics have been proposed \cite{xu2024engineering2dsquarelattice}. While the monolayer only has $C_{2v}$ symmetry, the $90^{\circ}$-twisted stacking generates an effective moiré square lattice model. The tunability of this promising setup, however, is limited as the angle is fixed and the moiré scale is not significantly larger than that of the parent lattice scale. Across various possible setups mentioned above, the twisted homobilayer stands out as the most minimalist and versatile choice.

To the end of simulating a single-orbital square lattice Hubbard model, the monolayer material should have its band extremum sitting at the Brillouin zone center ($\Gamma$ point) or corner ($M$ point). We will focus on the latter case as it is believed to have stronger interference effects \cite{PhysRevResearch.1.033076}, leading to deeper moiré potentials. Ref. \cite{Pavel} used a minimal theoretical model to examine this possibility and found a widely tunable $t-t'-U$ Hubbard model when the displacement field is present. While some monolayer square lattice materials have been fabricated in the lab (see for example \cite{ji2019freestanding,shigekawa2019dichotomy,yu2019high,Wang_2012}), simple non-degenerate $M$-point materials are sparse \cite{mounet2018two} and have not been experimentally realized to our best knowledge. 
One promising candidate is the monolayer carbon allotrope C$_{568}$, shown to be stable via first-principle calculations by independent groups \cite{RAM2020827,20050034,doi:10.1080/1536383X.2021.1944119,ZHAO2023154895}. Its geometry contains pentagon, hexagon and octagon plaquettes, therefore the name. The conduction band minimum is predicted to be at the $\Gamma$-point, while the valence band maximum is at the $M$-point. Although hosting a square lattice structure, the material is non-planar and buckled such that the $C_{4z}$ symmetry is absent. Instead, there is a $S_4$ symmetry which is the composition of $C_{4z}$ and reflection with respect to the monolayer plane. This distinguishes C$_{568}$ from the previous theoretical analysis in ref. \cite{Pavel}. 

In this work, we examine  twisted homobilayers of C$_{568}$. Through symmetry and first-principle analyses, we write down the bilayer continuum Hamiltonian with realistic parameters. Upon studying the band structures, both with and without external displacement fields, we derive the corresponding effective moiré Hubbard models. We find the effective model without field to be that of the single-band Hubbard model on the square lattice, unlike the case considered in Ref. \cite{Pavel}. Upon tuning the twisting angles, the nearest and next-nearest neighbor hopping strengths cross at around $8.5^{\circ}$. A widely tunable $t'/t$
ratios plays an important role
in realizing superconductivity in square lattice $t-t'-U$ Hubbard model \cite{doi:10.1126/science.aal5304,PhysRevB.102.041106,PhysRevResearch.2.033073,danilov2022degenerate} and $t-t'-J$ models \cite{PhysRevLett.132.066002,doi:10.1073/pnas.2109978118,doi:10.1073/pnas.2420963122}, and also leads to the potential of frustrated magnetism \cite{PhysRevB.86.024424,PhysRevX.11.031034,PhysRevB.109.L161103,PhysRevB.109.184410}. When displacement field is further added, an anisotropic nearest neighbor hopping $t_x\neq t_y$ can be reached, leading to imbalanced band flattening in the two directions, also found in moiré systems with generic symmetry, see for example \cite{PhysRevResearch.1.033076,PhysRevB.107.085127,PhysRevB.107.174506}. 
We estimate the magnitude of both the onsite and the longer ranged the Coulomb interactions. At smaller twisting angles and fields, the system is in the strongly interacting regime. For example, when $\theta = 3^{\circ}$ and $D=0$, we have the ratio $U/t_x\sim 10^5$, and $t_{x+y}/t_{x}=-0.7639$.  

We conclude the introduction by pointing out some of the other works on moiré square lattice systems which serve different purposes, including: twisted bilayers of uniform \cite{PhysRevB.105.165422} or staggered flux states \cite{PhysRevB.104.035136}, states with quadratic band touching \cite{PhysRevResearch.4.043151}, twisted bilayers of cuprates \cite{can2021high,doi:10.1126/science.abl8371,PhysRevB.105.L201102,PhysRevB.106.104505,PhysRevLett.130.186001} and FeSe \cite{10.21468/SciPostPhys.15.3.081}.

\section{Modeling from first-principles} 

\begin{figure*}
    \centering
    \includegraphics[]{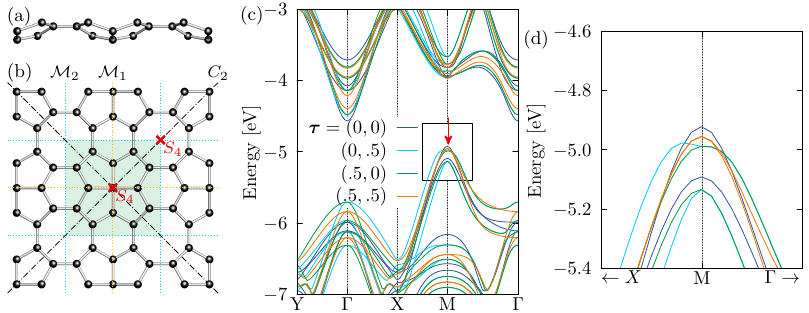}
    \caption{Side view (a) and the top view (b) of the lattice structure of monolayer C$_{568}$. In the top view, a unit cell is also indicated. The yellow and cyan dotted lines are for the reflections $\mathcal{M}_1$ and $\mathcal{M}_2$, respectively. The dash-dotted lines are for $C_2=\mathcal{M}_d\mathcal{M}_{xy}$. The red cross marks are for the $S_4$ centers. (c) Band structures of untwisted bilayers with several selected relative shifts $\bm{\tau}$. In the picture, $\bm{\tau}$ is indicated in the unit of $a_0$. (d) The close up view of the boxed area in (c).}
    \label{fig:lattice_and_band}
\end{figure*}
The lattice structure of monolayer C$_{568}$ is shown in Figs.~\ref{fig:lattice_and_band}(a) and \ref{fig:lattice_and_band}(b). %The name C$_{568}$ is after the fact that the lattice structure contains 5-, 6-, and 8-membered carbon rings. 
Interestingly, the network consists of mixture of sp$^3$ and sp$^2$ bondings, where the four sp$^3$ bonds locate at the center of unit cell in Fig.~\ref{fig:lattice_and_band}(b). 
Due to the sp$^3$ nature, the monolayer is buckled as clearly seen in Fig.~\ref{fig:lattice_and_band}(a). A key monolayer symmetry is roto-inversion $S_4$ associated with the sp$^3$ bonds, which can equivalently be expressed as $C_{4z}\mathcal{M}_{xy}$, where $\mathcal{M}_{xy}$ is a reflection with respect to the plane of monolayer. Note that $S_4$ automatically gives $C_{2z}$ with respect to the $S_4$ center. (Due to the lattice translation symmetry, there appears the other $S_4$ center at the corner of a unit cell.) The monolayer breaks inversion symmetry. The other important symmetry is reflection with respect to the  plane perpendicular to the layer and running along horizontal or vertical directions in Fig.~\ref{fig:lattice_and_band}(b). The plane passes through either the center ($\mathcal{M}_1$) or the corner  ($\mathcal{M}_2$) of a unit cell. There is also a $C_2$ axis along the diagonal direction in Fig.~\ref{fig:lattice_and_band}(b). This in-plane $C_2$ operation can be decomposed into $\mathcal{M}_{xy}$ and a reflection with respect to the diagonal lines in Fig.~\ref{fig:lattice_and_band}(b), which we name $\mathcal{M}_d$.

Monolayer C$_{568}$ has been predicted  \cite{RAM2020827} to be a semiconductor with its valence top at the $M$-point in the Brillouin zone. These crystalline and electronic structures are reproduced in our density functional theory (DFT) calculations, see Appendix \ref{app:DFT} for computational details. In this paper, we focus on electronic structures near the valence top. 

The moir\'e pattern in twisted bilayer stems from the position dependence of relative shift $\bm{\tau}$ between two layers. For small twist angles, a moir\'e system is \textit{locally} well approximated by untwisted bilayer with relative lateral shift $\bm{\tau}$, and investigating $\bm{\tau}$-dependence of band energy near the valence top in the untwisted bilayer gives information for constructing a moir\'e twisted bilayer model. Since monolayer C$_{568}$ breaks spatial inversion symmetry, there are two types of stacking patterns: one with two identical copies of monolayers and one between two inversion images of each other. The former still breaks inversion symmetry after stacking while the latter restores inversion symmetry after stacking. In this paper, we focus on the former with the aim of realizing the square lattice Hubbard model in twisted bilayers.

%Investigation of $\bm{\tau}$-dependence starts with deriving $\bm{\tau}$-dependence of the layer distance, which corresponds to corrugation in moir\'e bilayers. For this, we derive total energies within DFT for each $\bm{\tau}$ as functions of interlayer distance with fixed intralayer coordinates, and determine the optimized distance by minimizing the total energy. (See Appendix for details.) Using the optimized distance for each $\bm{\tau}$, the band structures are derived in a normal procedure in DFT. 
For different relative shifts $\bm{\tau}$, the interlayer distances $d_z (\bm{\tau})$ can be different. Therefore, such corrugation effect in the moi\'e bilayers must be taken into account when comparing the bilayer band structures for different $\bm{\tau}$'s. For a fixed stacking pattern $\bm{\tau}$, we calculate the corresponding total energy using DFT for various interlayer distances, and fix the optimal $d_z (\bm{\tau})$ to be the one that minimizes the energy. (See Appendix \ref{app:dz} for details.) 
Because of layer doubling, there are two bands near the valence top, and the energies of these two bands are used in extracting parameters for an effective model. Figure~\ref{fig:lattice_and_band}(c) shows bilayer band structures for several selected $\bm{\tau}$s. Focusing on the valence top at M-point, sizable $\bm{\tau}$ dependence is observed, indicative of strong band reconstruction in twisted bilayer cases. 

The effective continuum model for the untwisted bilayer expanded near the Brillouin zone corner $\mathbf{k}_M$ can be expanded in the layer basis as
\begin{equation}
    H= H_0(\bm{k}_M) \sigma_0 + V_{\text{pot}}(\bm{\tau}),\quad V_{\text{pot}}(\bm{\tau})=\sum_{a=0}^3 V_{a,\bm{k}_M} (\bm{\tau}) \sigma_a,
\label{eq:untwisted_continuum_model}
\end{equation}
where $\sigma_a$ labels the Pauli matrices acting in the layer space and $\sigma_0$ is identity. 
The functions $H_0$ and $V_{\text{pot}}$ both depend on the base momentum $\textbf{k}_M$ and are subject to symmetry constraints of the bilayer system. $V_{\text{pot}}$ further depends on the relative displacement between the two layers $\bm{\tau}$.

The associated energies are 
% \begin{equation}
%     E_{\pm}(\bm{k}_M,\bm{\tau}) = H_0(\bm{k}_M) + V^{(0)}_{\bm{k}_M}(\bm{\tau})
%     \pm \sqrt{V^{(3)}_{\bm{k}_M}(\bm{\tau})^2+|V_{\bm{k}_M}(\bm{\tau})|^2}.
% \end{equation}
\begin{equation}
    E_{\pm}(\bm{k}_M,\bm{\tau}) = H_0(\bm{k}_M) + V_{0,\bm{k}_M}(\bm{\tau})
    \pm \left(
    \sum_{a=1}^3
    V_{a,\bm{k}_M}(\bm{\tau})^2
    \right)^{1/2}.
\end{equation}
We choose the zero-energy level  $H_0(\bm{k}_M)=0$ for convenience and will omit the subscript $\bm{k}_M$ in $V_a$ from now on. The $V$-functions can be determined by comparing these energies with the DFT results.
% For constructing a moir\'e effective model where the Brillouin zone is much smaller than the original Brillouin zone, $\bm{k}$ close to the M-point is especially important. So, we introduce $V_i(\bm{\tau})$ as 
% \begin{equation}
%     V_0 = V^{(0)}_{\bm{k}_M},\,
%     V_1 = \mathrm{Re}V_{\bm{k}_M},\
%     V_2 = \mathrm{Im}V_{\bm{k}_M},\
%     V_3 = V^{(3)}_{\bm{k}_M},
% \end{equation}
% where $\bm{k}_M$ is the momentum at M-point and the argument $\bm{\tau}$ is suppressed for simplicity. 
%Later, we replace $H_0(\bm{k})$ by $H_0(-i\bm{\nabla})$, and for that, it is convenient to set $H_0(\bm{k}_M)=0$, which is merely a choice of zero energy level.
%Using $V_i$, the potential term $V_{\text{pot}}(\bm{\tau})$ in Hamiltonian, which simply consists of terms other than $H_0(\bm{k})$ term, can be written as 
% \begin{equation}
%     V_{\text{pot}}(\bm{\tau}) = \sum_iV_i\sigma_i,
% \end{equation}
% where $\sigma_i$ are Pauli matrices with $\sigma_0$ being an identity matrix. 

The symmetry of $V_{\text{pot}}$ is essential for constructing an effective model. In the monolayer case,  roto-inversion symmetry $S_4=C_{4z}\mathcal{M}_{xy}$. For bilayers, $\mathcal{M}_{xy}$ interchanges two layers, therefore the $S_4$ constraint reads 
\begin{equation}
    V_{\text{pot}}(C_{4z}\bm{\tau}) = \sigma_1V_{\text{pot}}(\bm{\tau})\sigma_1. \label{eq:V_symm_1}
\end{equation}
More specifically
% \begin{equation}
%     V_{0,1}(C_{4z}\bm{\tau})=V_{0,1}(\bm{\tau}),\,
%     V_{2,3}(C_{4z}\bm{\tau})=-V_{2,3}(\bm{\tau}). \label{eq:symm_const_1}
% \end{equation}
$V_{0,1}(C_{4z}\bm{\tau})=V_{0,1}(\bm{\tau})$ and   
$V_{2,3}(C_{4z}\bm{\tau})=-V_{2,3}(\bm{\tau}).$
The monolayer in-plane $C_2=\mathcal{M}_d\mathcal{M}_{xy}$, and leads to 
\begin{equation}
    V_{\text{pot}}(\mathcal{M}_d\bm{\tau}) = \sigma_1V_{\text{pot}}(\bm{\tau})\sigma_1. \label{eq:V_symm_2}
\end{equation}
Specifically this gives constraints
$V_{0,1}(\mathcal{M}_d\bm{\tau})=V_{0,1}(\bm{\tau})$
and $V_{2,3}(\mathcal{M}_d\bm{\tau})=-V_{2,3}(\bm{\tau}). $
% \begin{equation}
%     V_{0,1}(\mathcal{M}_d\bm{\tau})=V_{0,1}(\bm{\tau}),\,
%     V_{2,3}(\mathcal{M}_d\bm{\tau})=-V_{2,3}(\bm{\tau}). \label{eq:symm_const_2}
% \end{equation}
Lastly, since the M-point is  a high symmetry point in the Brillouin zone,  quantum interference can eliminate interlayer tunneling for specific $\bm{\tau}$ points \cite{PhysRevB.95.245401,PhysRevResearch.1.033076}. In this case, the reflection symmetries $\mathcal{M}_1$ and $\mathcal{M}_2$ in fig. \ref{fig:lattice_and_band} in monolayer lead to $V_{1,2}(\bm{\tau})=0$ when $\bm{\tau}$ is at the boundary of a unit cell. 
In addition, on the diagonal lines in the unit cell which satisfy $\mathcal{M}_d\bm{\tau}=\bm{\tau}$, $V_3$ vanishes. Also, since M-point is time reversal invariant momentum, there exists a basis such that the Hamiltonian becomes real. In that basis, $V_2$ is zero and there remain three functions to be determined. We remark that if the inversion symmetry was present, then the $V_3(\bm{\tau})$ term would not have been allowed, which was the case in \cite{Pavel}.

However, easily obtained inputs from DFT are only two functions, $E_{\pm}(\bm{k}_M,\bm{\tau})$. To overcome this issue, we adopt lower harmonics expansion of $V_i$ respecting crystalline symmetries, and fix required coefficients by the least square fitting to DFT data. (See Appendix \ref{app:harmonics} for details.) The prepared lower harmonics expansions are
% \begin{align}
%     V_0(\bm{\tau}) =& \gamma_0+\gamma_1(\cos\tilde{\tau}_x+\cos\tilde{\tau}_y)\\
%     &+\gamma_2(\cos(\tilde{\tau}_x+\tilde{\tau}_y)+\cos(\tilde{\tau}_x-\tilde{\tau}_y))\\
%     &+\gamma_3(\cos 2\tilde{\tau}_x+\cos 2\tilde{\tau}_y)\\
%     &+\gamma_4(\cos(2\tilde{\tau}_x+\tilde{\tau}_y)+\cos(\tilde{\tau}_x+2\tilde{\tau}_y)\\
%     &\quad\quad +\cos(2\tilde{\tau}_x-\tilde{\tau}_y)+\cos(\tilde{\tau}_x-2\tilde{\tau}_y)),\\
%     V_1(\bm{\tau}) =& \alpha_1\cos\frac{\tilde{\tau}_x}{2}\cos\frac{\tilde{\tau}_y}{2}\\
%     &+\alpha_2(\cos\frac{3\tilde{\tau}_x}{2}\cos\frac{\tilde{\tau}_y}{2}+\cos\frac{\tilde{\tau}_x}{2}\cos\frac{3\tilde{\tau}_y}{2}),\\
%     V_3(\bm{\tau}) =& \beta_1(\cos\tilde{\tau}_x-\cos\tilde{\tau}_y)+\beta_2(\cos 2\tilde{\tau}_x-\cos 2\tilde{\tau}_y)\\
%     &+\beta_3(\cos 3\tilde{\tau}_x-\cos 3\tilde{\tau}_y),
% \end{align}
\be
\begin{split}
    V_0(\bm{\tau}) = &\ \gamma_0+\gamma_1(\cos\tilde{\tau}_x+\cos\tilde{\tau}_y)+2\gamma_2\cos(\tilde{\tau}_x)\cos(\tilde{\tau}_y)\\
    &+\gamma_3(\cos 2\tilde{\tau}_x+\cos 2\tilde{\tau}_y)\\    &+2\gamma_4(\cos(2\tilde{\tau}_x)\cos(\tilde{\tau}_y)+\cos(\tilde{\tau}_x)\cos(2\tilde{\tau}_y)),\\
    V_1(\bm{\tau}) =& \alpha_1\cos\frac{\tilde{\tau}_x}{2}\cos\frac{\tilde{\tau}_y}{2}\\
    &+\alpha_2\big(\cos\frac{3\tilde{\tau}_x}{2}\cos\frac{\tilde{\tau}_y}{2}+\cos\frac{\tilde{\tau}_x}{2}\cos\frac{3\tilde{\tau}_y}{2}\big),\\
    V_3(\bm{\tau}) =& \beta_1(\cos\tilde{\tau}_x-\cos\tilde{\tau}_y)+\beta_2(\cos 2\tilde{\tau}_x-\cos 2\tilde{\tau}_y)\\
    &+\beta_3(\cos 3\tilde{\tau}_x-\cos 3\tilde{\tau}_y),
\end{split}
\ee
where $\tilde{\tau}_\mu=2\pi\tau_\mu/a_0$. These expansions satisfy all the symmetry constraints described above. %the constraints Eqs.~\ref{eq:V_symm_1} and \ref{eq:V_symm_2}, and the requirement for $V_1$ vanishes at the boundary of a unit cell. 

\begin{figure}
    \centering
    \includegraphics[]{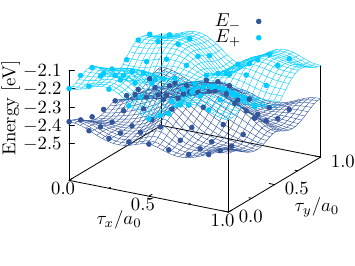}
    \caption{$E_+(\bm{k}_M,\bm{\tau})$ and $E_-(\bm{k}_M,\bm{\tau})$ obtained with the lower harmonics expansion (lines) and DFT (points). }
    \label{fig:map_disp_with_corrugation}
\end{figure}

When fitting with the DFT data, coefficients in $V_0$ can be determined using a data set $(E_++E_-)/2$. The remaining data set $(E_+-E_-)/2$ can be used for determining two functions $V_1$ and $V_3$: at the boundary of a unit cell, $V_1=0$ and $(E_+-E_-)/2$ is solely determined from $V_3$; on the other hand, on the diagonal lines in a unit cell that satisfies $\mathcal{M}_{d}\bm{\tau}=\bm{\tau}$, the symmetry constraints leads to $V_3=0$, and $(E_+-E_-)/2$ is solely determined from $V_1$. 
Using these facts, we obtain
\be
\begin{split}
& \alpha_1 = 0.194,\ \alpha_2 = -0.052,\ \gamma_0 = -2.292,\\
& \gamma_1 = -0.011,\ \gamma_2=0.021,\ 
    \gamma_3 = -0.007,\ \gamma_4 = -0.001,\\
& \beta_1 = -0.024,\ \beta_2 = -0.008,\ \beta_3= -0.011,
\end{split}
\label{eq:parameters}
\ee
% \be
% \begin{split}
% & (\alpha_1,\alpha_2,\beta_1,\beta_2,\beta_3)=(0.194,-0.052, -0.024,-0.008,-0.011)\\
% & (\gamma_0,\gamma_1,\gamma_2,\gamma_3,\gamma_4)=(-2.292,-0.011,0.021,-0.007,-0.001).
% \end{split}
% \ee
all in the unit of eV.  The lower harmonics expansion and the DFT results are compared in Fig.~\ref{fig:map_disp_with_corrugation}. Considering the simplicity of the approximation and limitation in converting $E_{\pm}$ data set to the coefficients, the match between the data and the lower harmonics expansion is remarkable.

\section{Twisted bilayer model}  

\begin{figure}[htbp]
\centering
\includegraphics{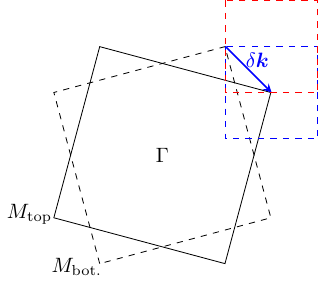}
\includegraphics[scale=0.8]{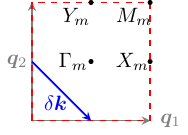}
\caption{Left: Brillouin zones of the top and bottom layers. Dashed blue square is the moiré Brillouin zone, and dashed red square is the shifted moiré Brillouin zone under the transformation $U$. 
Right: High symmetry points in the shifted moiré Brillouin zone. $\Gamma_m$ point has momentum $\bm{k}_0.$}
\label{fig:BZ}
\end{figure}

In this section, we specify to the rigid twist case with dispacement $\bm{\tau}(\bm{r}) = 2\sin\frac{\theta}{2}\hat{z}\times\bm{r}.$ The moiré lattice constant is thus $a_M\approx a_0/\theta$ at small twisting angles. We will denote the moiré lattice vectors by $\bm{R}_1, \bm{R}_2$ and the moiré reciprocal lattice vectors by $\bm{q}_1, \bm{q}_2$. The continuum Hamiltonian reads
\begin{equation}
    H=
    \begin{pmatrix}
    H_-(-i\bm{\nabla})+V_0+V_3 & e^{\mathrm{i}\bm{\delta k}\cdot\bm{r}}V_1\\
    e^{-\mathrm{i}\bm{\delta k}\cdot\bm{r}}V_1 &  H_+(-i\bm{\nabla})+V_0-V_3
    \end{pmatrix},
\label{eq:model}
\end{equation} 
where $H_{\pm}(-i\bm{\nabla})=\frac{\hbar^2}{2m_*}(-i\bm{\nabla}-\bm{k}_M\pm\delta\bm{k}/2)^2$ and %$\delta\bm{k}=\frac{\pi}{a_M} (-1,1)$.
$\delta\bm{k}=2\sin \frac{\theta}{2} ( \hat{\bm{z}}\times \bm{k}_M)=(\bm{q}_1-\bm{q}_2)/2,$ which is depicted in Fig. \ref{fig:BZ}.
%$\mu=\frac{\hbar^2}{2m_*}$

When $V_0=V_3=0$ and in \eqref{eq:model} and keeping only the leading term in $V_1$, i.e., $V_1=\alpha_1 \cos(\frac{\pi}{a_0} \tau_1)\cos(\frac{\pi}{a_0} \tau_2)=\alpha_2\cos (\frac{\pi}{a_M} y)\cos (\frac{\pi}{a_M} x),$ the model reduces to that studied in \cite{Pavel}: there is an emergent symmetry at small twisting angles $\mathcal{L}_0=\sigma_x e^{i\sigma_z \delta \k\cdot \bm{r}}$. This symmetry forces the band maxima to be degenerate $E_{\Gamma}=E_M$ leads to the vanishing of nearest neighbor hopping in the effective tight binding model defined on the moiré lattice. %and the fact that $V_1(\bm{r}+\bm{R}_1)=-V_1(\bm{r})$ leads to two decoupled . 
Adding $V_0$ as well as the sub-leading terms in $V_1$ don't change the physics.

In C$_{568}$, inversion symmetry is broken, such that a finite $V_3$ term is allowed which breaks $\mathcal{L}_0$. However, a different emergent symmetry, which involves the fourfold rotation, becomes present at small twisting angles. To understand the effect of the emergent symmetry on the moiré momentum space, we apply the following convenient unitary transformation 
${\mathcal{U}}=e^{i\bm{k}_0\cdot \bm{r}} \sigma_0 $ where $\bm{k}_0=\bm{k}_M+(\bm{q}_1+\bm{q}_2)/4$
%$\bm{Q}=\frac{\pi}{a_M}(0,-1)$
, such that ${H}'={\mathcal{U}^{\dagger}} H {\mathcal{U}}$ has similar form as in \eqref{eq:model} but with $H_{\pm}$ replaced by $H_{\pm}'=  \frac{\hbar^2}{2m_*}(-i\bm{\nabla}-\bm{k}_{\bm{M}}+\bm{k}_0\pm \delta \bm{k}/2)^2$. We can now label any momentum using its shift from $\bm{k}_0:$  $\bm{q}=\bm{k}-\bm{k}_0$, where $\bm{k}_0$ is the $\Gamma_m$ point in the shifted Brillouin zone plotted in the right panel of figure \ref{fig:BZ}. Then we conveniently have 
\be
\begin{split}
& H_{+}'= \frac{\hbar^2}{2m_*}(-i\bm{\nabla}+\bm{q}+\bm{q}_{1}/2)^2,\\
& H_-'= \frac{\hbar^2}{2m_*}(-i\bm{\nabla}+\bm{q}+\bm{q}_2/2)^2. 
\label{eq:gauge_choice}
\end{split}
\ee
It is easy to check (details are shown in appendix \ref{app:symmetry}) that the following operation leaves the Hamiltonian invariant:
\be
\mathcal{L}=e^{-\frac{\mathrm{i}}{2}\bm{q}_2\cdot \bm{r} (\sigma_3+\sigma_0)/2} \sigma_1 C_{4z},\quad H'(\bm{q})=\mathcal{L} H'(R_{\pi/2} \bm{q}) \mathcal{L}^{-1}.
\label{eq:emergent}
\ee

Consequently, as long as the band is non-degenerate, the energy of the Hamiltonian $H'$ must satisfy
\be
E(\bm{q})=E(R_{\pi/2} \bm{q}),
\label{eq:degeneracy}
\ee
which leads to $E(X_m)=E(Y_m)$ in the shifted moiré Brillouin zone shown in fig. \ref{fig:BZ}.

\begin{figure}[htbp]
\flushleft
\includegraphics[scale=0.32]{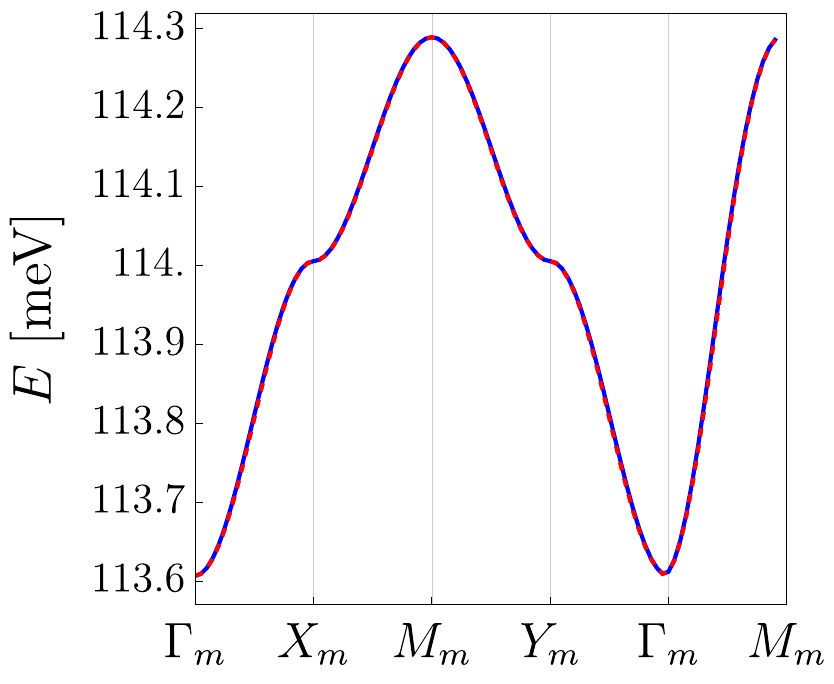}
\includegraphics[scale=0.309,trim={1.1cm 0 0 0},clip]{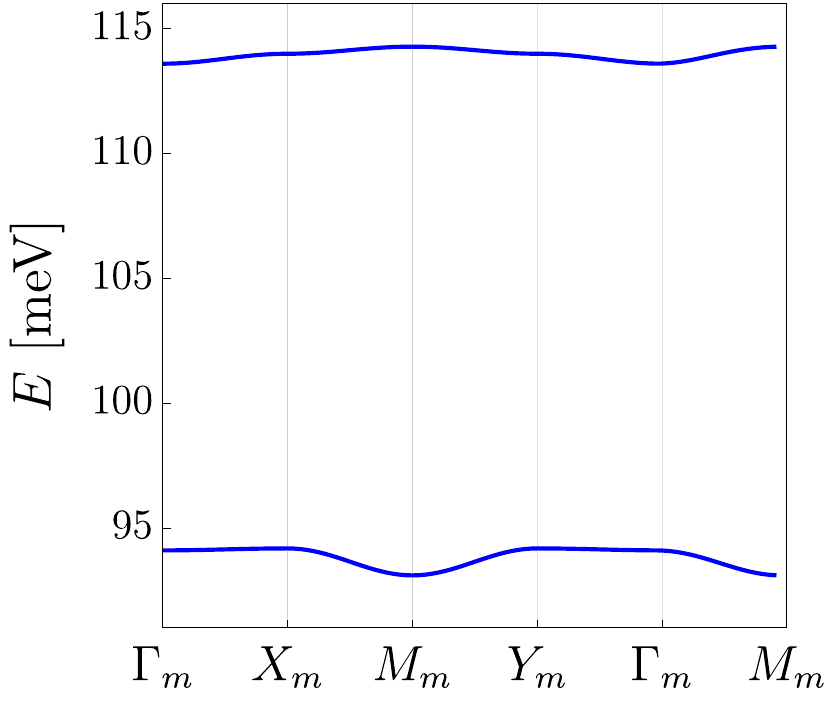}
\caption{Left: The top moiré band of \eqref{eq:model} at $\theta=3.6^{\circ}$ (blue). This clearly resembles the dispersion of tight binding model on square lattice (red dashed). Right: the next band is well-separated. }
\label{fig:band}
\end{figure}

Using the fitted parameters found in \eqref{eq:parameters} and keeping only those whose magnitudes are larger than $0.01$ eV , we show the band structure $E(\bm{q})$ along high symmetry lines in fig. \ref{fig:band}: the band top lives at the $M_m$ point, and the band bottom is at the $\Gamma_m$ point in the shifted Brillouin zone. The shoulders at $X_m$ and $Y_m$ points have the same energies consistent with \eqref{eq:degeneracy}. The band structure \ref{fig:band} clearly resembles that of the tight-binding Hamiltonian on square lattice with nearest neighbor hopping, which we will explain in more details in section \ref{subsec:hop}. This is in contrast to the case in ref. \cite{Pavel} where an external field is necessary for reaching the same dispersion. 
At $\theta=3.6^{\circ}$, the flat moiré band is separated from the neighboring band by a large gap of $\sim 19.5$ meV, which is about $28$ times the bandwidth of the top flat band.

\subsection{Displacement field}

\begin{figure}
\flushleft
\includegraphics[scale=0.34]{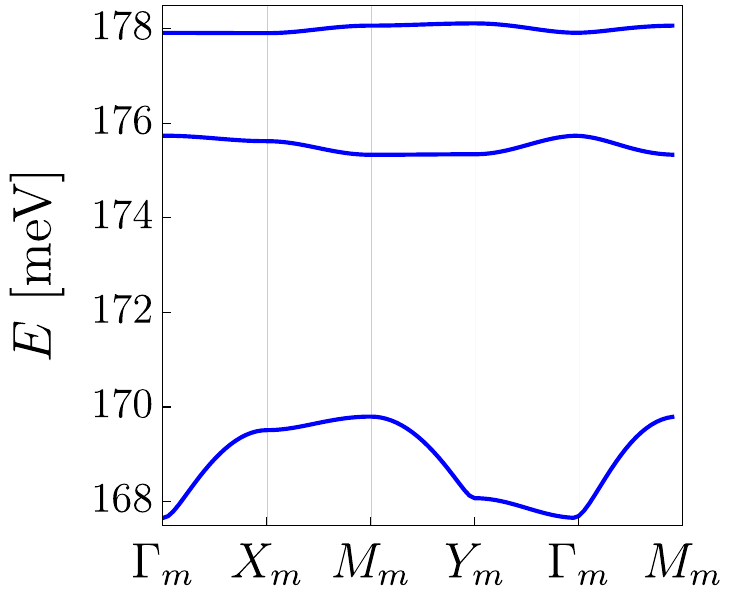}
\includegraphics[scale=0.34]{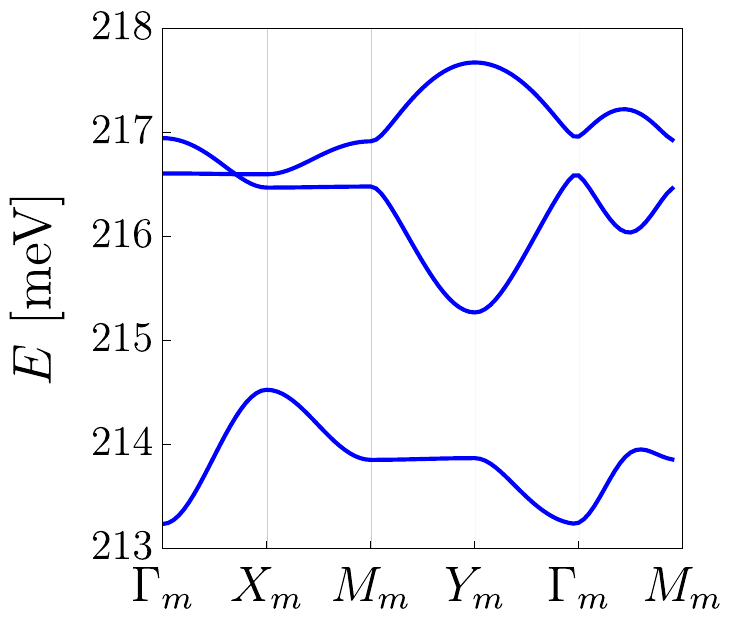}
\caption{Band structures at different values of displacement fields. Left: the top two moiré bands are well-separated at $D=-120$ meV. Right: the top two moiré bands cross at the $\Gamma-X$ high symmetry line when $D=-170$ meV, but are still well separated from other bands.}
\label{fig:D}
\end{figure}

Upon adding a displacement field $D\sigma_z,$ the emergent $\mathcal{L}$ symmetry which involves layer exchange is absent. In the shifted moiré Brillouin zone plotted in the right panel of fig. \ref{fig:BZ}, this translates into the asymmetry $E_X \neq E_Y.$ Taking into account all the terms in the Hamiltonian, a small and negative $D$ already shifts the band extrema: $E_Y>E_M>E_{\Gamma}>E_X$. 
Further increasing the magnitude, when  $D=-0.17$ eV, band crossings between the top two moiré bands are induced, and beyond which $E_M<E_{\Gamma}$. The band crossings always happen along the Brillouin zone boundary which respects the reflection symmetry. 
See Fig. \ref{fig:D} for typical band structures. For the a positive $D$, the evolution of band structures is similar but with $X\leftrightarrow Y$ exchanged.

\section{Tight binding model}

In this section we analyze the effective tight-binding model for the twisted bilayer. At small twisting angles, it is sufficient to include only the nearest neighbor $t_x, t_y$ 
and the next nearest neighbor hoppings $t_{x+y}$, $t_{x-y}$. Time reversal symmetry requires them to be real. 
Reflection symmetries with respect to the $x$- and $y$- axes require %the nearest neighbor hoppings $t_{\hat{\bm{x}}}, t_{\hat{\bm{y}}}$ to be real, and for 
the next nearest neighbor hoppings, %$t_{\hat{\bm{x}}+\hat{\bm{y}}}=t_{\hat{\bm{x}}-\hat{\bm{y}}}\equiv t'$, 
$t_{x+y}=t_{x-y}$. 
The corresponding dispersion is: 
% \be
% \begin{split}
% E(\bm{k})= & 
% E_0-t_x \cos k_x - t_y \cos k_y 
% \\ & 
% - t'_{x+y} \cos (k_x + k_y) - t'_{x-y} \cos (k_x-k_y).
% \end{split}
% \ee
\be
E(\bm{k})= E_0-t_x \cos k_x - t_y \cos k_y - 2t_{x+y} \cos k_x\cos k_y.
\label{eq:tight_energy}
\ee
The emergent $\L$ symmetry, when present, imposes that the nearest neighbor hopping $t_x=t_y$.

\subsection{The hopping parameters}
\label{subsec:hop}

The four parameters $\{E_0, t_x, t_y, t_{x+y}\}$ can be easily solved by requiring that at high symmetry points $\Gamma_m, X_m, Y_m, M_m$ defined in fig. \ref{fig:BZ}, the energies computed from the twisted bilayer model and that from the tight binding model \eqref{eq:tight_energy} should match with each other. This method works well at small twisting angles where the longer ranged hoppings can be ignored.  For example 
at $D=0$ and $\theta=3.6^{\circ}$, we have $t_x=t_y=0.171$ meV, $t_1=0.014$ meV. We plot the dispersions calculated both from the tight-binding model and the lowest-harmonics twisted bilayer model in fig. \ref{fig:band}(a) and they match nicely. 
While one can also plot $t$ as a function of twisting angle and displacement field using this simple method, below we will perform a more careful numerical examination. 
\begin{figure}
    \centering
    \includegraphics[]{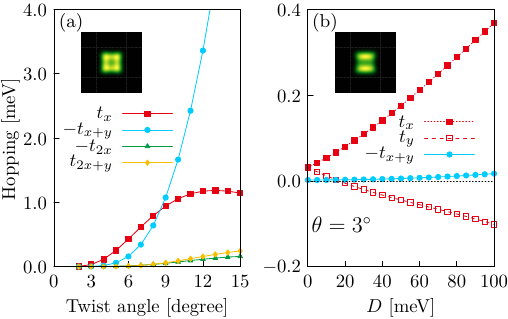}
    \caption{Hopping parameters derived using full parameters and  Wannier functions. (a) Twist angle dependence. $t_x$, $t_{x+y}$, $t_{2x}$, and $t_{2x+y}$ correspond to the first to the fourth nearest neighbor hoppings in the square lattice. (b) Displacement field dependence for $\theta=3^\circ$. Insets: total charge (sum of the charges on the upper and the lower layer) for a Wannier function at $\phi=3^\circ$ with (a) $D=0$ and (b) $D=50$ meV. }
    \label{fig:hoppings_wannier}
\end{figure}
%So far, we focused on a simplified model to see essential physics in twisted bilayer C$_{\text{568}}$ and to compare with the former theory on M-point materials. To see the angle dependence of parameters in the effective tight-binding model, we elaborate our numerics in two aspects: 
The elaboration is done in the following aspects: 
(i) taking all $\gamma_i$, $\alpha_i$, and $\beta_i$ into account, and (ii) deriving the Wannier orbitals for the highest energy band. In the parameter range showing no  crossing between the top and  second bands, the highest energy band is well isolated from the other bands, and it is justified to have reasonably localized Wannier orbitals. In practice, we prepare Bloch wave functions for the highest energy band in the continuum Hamiltonian Eq.~\eqref{eq:model} (with full account of $\gamma_i$, $\alpha_i$, and $\beta_i$) on $13\times 13$ regular grid in the moire Brillouin zone, and project them to an ansatz Gaussian function to fix the phases required in constructing Wannier functions. (This procedure is the same as the projection on initial guess functions in Ref.~\cite{PhysRevB.56.12847}.) 

The obtained hopping parameters are plotted in Fig.~\ref{fig:hoppings_wannier}. The angle dependence without displacement field is summarized in Fig.~\ref{fig:hoppings_wannier}(a). A characteristic crossover from the regime $|t_x|<|t_{x+y}|$ to the regime $|t_x|>|t_{x+y}|$ takes place at around $8.5^\circ$ as the twist angle is decreased. Within the investigated range of the twist angle, $|t_{2x}|$ and $|t_{2x+y}|$ are significantly smaller than $|t_x|$ and $|t_{x+y}|$,
indicating the validity of the model Eq.~\eqref{eq:tight_energy}. The displacement field $D$ dependence at $\theta=3^\circ$ is shown in Fig.~\ref{fig:hoppings_wannier}(b). As we have seen in the simplified model, $t_x=t_y$ for $D=0$ due to the emergent $\mathcal{L}$ symmetry, and finite $D$ lifts this degeneracy. Interestingly, there is a point where $t_y=0$, i.e., almost purely one-dimensional state. (There remains slight two-dimensional nature due to small but finite $|t_{x+y}|$ and longer range hoppings.) In the large $D$ side, $|t_x|$ and $|t_y|$ grows much faster than $|t_{x+y}|$. 

The insets in Fig.~\ref{fig:hoppings_wannier} show the typical Wannier functions where the color corresponds to the sum of the wave function weights for the upper and the lower layers, which we write $\rho(\bm{r})$. The $S_4$ symmetry involves the layer exchange, but if we focus on the sum over the two layers, $\rho(\bm{r})$ should look $C_4$ symmetric for $D=0$, which is consistent with the inset of Fig.~\ref{fig:hoppings_wannier}(a). On the other hand, finite $D$ breaks the $S_4$ symmetry, $\rho(\bm{r})$ should look distorted, which is consistent with the inset of Fig.~\ref{fig:hoppings_wannier}(b). 

\subsection{Hubbard $U$ parameters}

Having the Wannier functions, the matrix elements for Coulomb interaction, in other words the Hubbard $U_{\bm{R}}$ parameters can be estimated, where $\bm{R}$ represents relative coordinate between two Wannier functions. See for example reference \cite{PhysRevB.75.224408} for the detailed method. We have adapted the formula
\begin{equation}
\begin{split}
    U_{\bm{R}} &= \int d^2\bm{r} \rho(\bm{r}+\bm{R})V(\bm{r}-\bm{r}')\rho(\bm{r}')\\
    &= \int \frac{d^2\bm{q}}{(2\pi)^2}\rho_{\bm{q}}\rho_{-\bm{q}}V_{\bm{q}}e^{-i\bm{q}\cdot\bm{R}},    
\end{split}
 \label{eq:U_estimation}
\end{equation}
where $\rho_{\bm{q}}$ is a Fourier component of $\rho(\bm{r})$, and
\begin{equation}
    V_{\bm{q}} = \frac{e^2}{2\epsilon\epsilon_0|\bm{q}|}\tanh (|\bm{q}|\xi).
        \label{eq:Coulomb_Vq}
\end{equation}
Here, we assume that the Coulomb interaction is screened by top and bottom metallic gates placed with the distance $\xi$ from the sample \cite{PhysRevB.100.205113,PhysRevB.102.035161,PhysRevB.102.045107}. From Eq.~\eqref{eq:Coulomb_Vq}, we can see that the screening effect is significant when the length scale $1/|\bm{q}|$ is much larger than $\xi$. Note that this is an estimation of the upper limit of $U$ because the screening effect from higher energy bands in addition to the one from the substrate has been ignored. 

\begin{figure}
    \centering
    \includegraphics[]{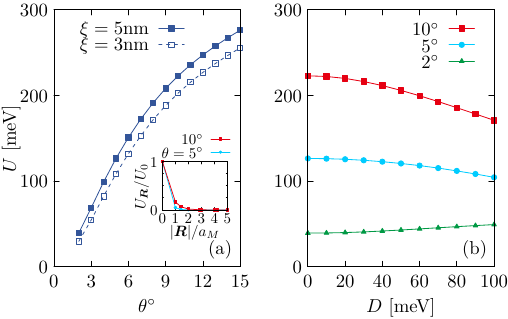}
    \caption{Hubbard $U$ estimated using the Wannier functions. (a) Twist angle dependence for $\xi=3$nm and $\xi=5$nm. Inset shows $|\bm{R}|$ dependence at $\xi=3$nm. (b) Displacement field dependence for selected twisted angles with $\xi=5$nm.}
    \label{fig:CoulombU_angle_disp}
\end{figure}
The twist angle dependence of $U\equiv U_0$ without displacement field is plotted in Fig.~\ref{fig:CoulombU_angle_disp}(a). Here, we employ $\epsilon=6$, a typical value for a hBN substrate, to take into account the screening effect from the substrate. The Hubbard $U$ is typically in the range of a few hundreds of meV, which is significantly larger than the hopping parameters. As the twist angle gets smaller, $U$ is also decreasing since the moir\'e length scale becomes larger. Figure~\ref{fig:CoulombU_angle_disp}(b) is for the displacement field dependence, showing that $U$ only weakly or moderately depends on $D$. The inset of Fig.~\ref{fig:CoulombU_angle_disp}(a) is for $|\bm{R}|$ dependence of $U_{\bm{R}}$ with $\xi=3$nm, showing rapid decay of $U_{\bm{R}}$. The decay rate is determined by the screening effect from the metallic gates, and depends on the twist angle, since the length scale is different for different twist angles.

In order to have an intuition on the size of $U$, we estimate the Coulomb interaction via $U(\epsilon)=\frac{e^2}{4\pi \epsilon \epsilon_0 l}$, where $l$ is the size of the Wannier orbital. The latter can be obtained from the harmonic oscillator approximation of the leading moiré potential $\alpha_1:$
\be
\frac{\hbar^2}{2m_*}\frac{1}{l^2}=\frac{1}{2}\alpha_1 \pi^2\frac{l^2}{a_M^2}.
\ee
Inserting  realistic numbers $m_*=0.64m_0,$ $a_0=5.725$\r{A}, we can solve $l=1.6515$ nm at $\theta=3^{\circ}$ and $D=0$. If the relative permittivity is taken to be $\epsilon=6,$ it leads to $U=0.1453$ eV, which is significantly larger than the bandwidth of $0.2984$ meV for the same parameter choice. If instead one plugs in the $l$ found from Wannier function calculations described above, at $\theta=3^{\circ}$ we would have $l=2.9562$ nm and $U=0.0812$ eV. These estimates match with the orders of magnitude in first-principle calculation, but the detailed discrepancies could be due to the oversimplified treatment where we only take into account $\alpha_1$ and the spherical approximation of the potential.

%\subsection{Possible Metal-insulator transition}

%Comparing the onsite Coulomb potential with the bandwidth of the top moiré band, we find that a potential metal-insulator transition can happen 

\section{Discussion}
We proposed twisted homobilayer C$_{568}$ to be a promising device in realizing the (extended) single-band square-lattice Hubbard model. If successfully fabricated, the device could shed important light on the mechanism of high temperature superconductivity realized in other square lattice materials. 
%could be an important stepping stone towards understanding and harnessing high-temperature superconductivity. 
At negligible displacement field, the system is naturally in the strong coupling limit, exhibiting a remarkably high $U/t$ ratio. For instance, when the twisting angle is at $3^{\circ}$ and $9^{\circ}$, the $U/t$ ratio is of the order $10^5$ and $10^2$, respectively. One can then probe the intermediate temperature regime,  of $t\ll T\ll U$, which is hard to achieve in  electronic materials.  Taking into account additional screening effects from higher energy bands, we expect these  $U/t$ values to decrease, potentially reaching a minimum value of $U/t\sim 10$, a regime closely resembling cuprate physics.
Additionally, the bilayer offers great flexibility: by tuning the twisting angle, a wide range of ratios between next-nearest-neighbor and nearest-neighbor hoppings can be achieved. Furthermore, the introduction of a displacement field induces anisotropy between hoppings in different directions $t_x, t_y$. The orientation of this anisotropy can be rotated by $90^\circ$ simply by reversing the sign of the vertical electric field. 
Notably, within an experimentally accessible range, the hopping amplitudes can change sign upon increasing $|D|$, with $t_{y}$ $(t_x)$ crossing from positive to negative when $D>0$ $(D<0)$.

\begin{figure}[htbp]
\centering
\includegraphics[scale=0.41]{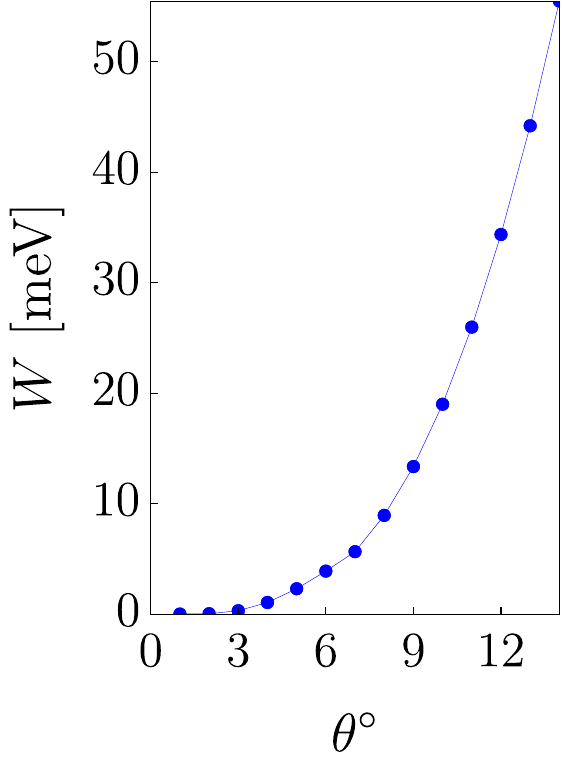}
\ \
\includegraphics[scale=0.41]{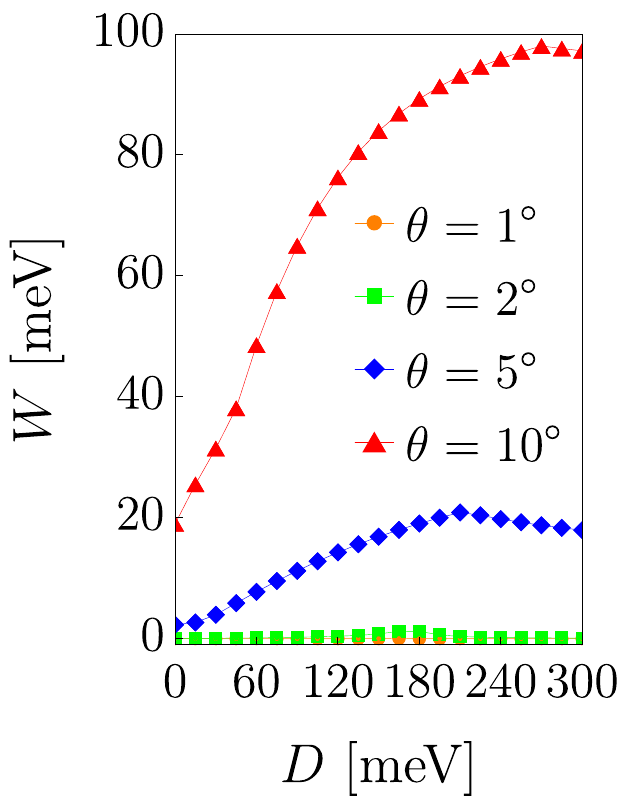}
\caption{Left: Band width $W$ as a function of twisting angle when displacement field $D=0$. Right: Band width as a function of $D$ for different twisting angles.}
\label{fig:width}
\end{figure}

In addition to reproducing the target single-band square lattice Hubbard model, twisted homobilayer C$_{568}$ can host interesting new phenomena. One possibility is tunable metal-insulator transition. 
Fig. \ref{fig:width} plots the bandwidth of the top moiré band. Away from band crossing, the bandwidth reflects the magnitudes of the hopping amplitudes in the system. Comparing with the onsite Coulomb energy shown in Fig. \ref{fig:CoulombU_angle_disp}, we see that at larger twisting angles and larger displacement fields (such that the system is highly anisotropic), the Coulomb interaction  can have the same order of magnitude as that of the bandwidth. Consequently, there can potentially be a metal-insulator transition tuned by displacement field. However, we have also shown in Fig. \ref{fig:D} that the top two moiré bands will cross at larger displacement fields $|D|\sim 170$ meV. Therefore, a full description of the potential transition requires a two-band tight-binding model. Since the large parameter space is beyond the scope of the current paper, we leave a careful examination for future work. 

Another interesting variation of the model is the following. In this paper, we focus on the stacking of two identical copies of a monolayer. In the case of the other type of stacking, i.e., stacking between inversion images, both of Eqs.~\eqref{eq:V_symm_1} and \eqref{eq:V_symm_2} reduce to 
\begin{gather}
    V_{\text{pot}}(\bm{\tau}) = \sigma_1V_{\text{pot}}(\bm{\tau})\sigma_1,
\end{gather}
which kills $V_{2,3}$ and induces no constraint on $V_{0,1}$. When there is no constraint, there is no reason for low energy effective model to have the symmetry of the square lattice. Namely, we expect $|t_x|\neq |t_y|$ even in the absence of a displacement fields. Thus, the successful fabrication of this family of structures would both shed important light on old problems and stimulate a host of new investigations.

Finally, we briefly discuss the potential for experimental realization of this proposed twisted homobilayer system. Theoretical predictions \cite{RAM2020827} indicate that  C$_{568}$ not only exhibits stability at room temperature but also possesses a formation energy of  $423$ meV/atom, which is considerably lower than that of other prominent 2D carbon allotropes such as graphdyine ($823$ meV/atom), graphyne ($675$ meV/atom), and pentagraphene ($968$ meV/atom). Notably, even graphdiyne with high formation energy has already been experimentally synthesized on a copper surface through a cross-coupling reaction involving hexaethynylbenzene  \cite{B922733D}. This precedent makes it highly reasonable to anticipate the successful synthesis of C$_{568}$. Moreover, in light of the promising potential applications in high-performance nanodevices and as anodes in lithium-ion batteries (see, for instance, \cite{ZHAO2023154895}), the authors maintain an optimistic outlook regarding the experimental fabrication of C$_{568}$.

\section*{Acknowledgements}
Z.-X. L. thanks Trithep Devakul, Eslam Khalaf, Philip Kim and Jedediah  Pixley for enlightening conversations and Pavel Volkov and Paul Eugenio for collaborations on a related project. This research is partially supported by the
Simons Collaboration on Ultra Quantum Matter which
is a grant from the Simons Foundation (651440, Z.-X.
L.), and JSPS KAKENHI Grant Number JP24K06968 (T.K.). This work was  performed in part at the Aspen Center for Physics, which is supported by National Science Foundation grant PHY-2210452. The part of calculations in this study have been done using the Numerical Materials Simulator at NIMS, and the facilities of the Supercomputer Center, the Institute for Solid State Physics, the University of Tokyo.

\appendix
\section{Density functional theory}
\label{app:DFT}
In this study, the density functional theory calculations have been done using Quantum Espresso package \cite{Giannozzi_2009,Giannozzi_2017}. For the exchange-correlation functional, we have adopted 
rev-vdW-DF2 functional \cite{PhysRevB.89.121103} to take van der Waals forces into account. Pseudopotentials are from PSlibrary \cite{DALCORSO2014337,pslibrary}, and we use projector augmented wave method to include core state information in pseudopotentials. In order to handle our target two-dimensional material, a 30{\AA} thick supercell in $z$-direction is used. The parameters for our continuum model are derived with a naive supercell method, but we have confirmed that dipole field effects are small and our conclusions are not affected.

\section{Lower harmonics expansion and the least square fit}
\label{app:harmonics}
Let us assume $\bm{\tau}$ dependence of some physical quantity $Q(\bm{\tau})$ can be written as
\begin{equation}
    Q(\bm{\tau}) = \sum_{l=1}^{N}c_lg_l(\bm{\tau})
\end{equation}
where each $g_l(\bm{\tau})$ function is some combination of harmonic functions that respects symmetry and periodicity of the system. Here, $N$ is the number of coefficients, or in other words, the number of degrees of freedom. On the other hand, let us assume that we have data (from DFT) for that quantity $D(\bm{\tau})$ on some finite grid in $\bm{\tau}$ space. Basically, $c_l$ is simply fixed by minimizing
\begin{equation}
    \Xi = \sum_{\bm{\tau}\in\text{grid}}|Q(\bm{\tau})-D(\bm{\tau})|^2.
\end{equation}
However, we sometimes want to apply constraints. For instance, we may take a special care about the values at $\bm{\tau}$ with high symmetry, e.g., $\bm{\tau}$ at zone corners or zone boundaries. Then, $\Xi$ is minimized under the condition that $Q(\bm{\tau})$ values at selected $\bm{\tau}$ points are pinned down to $D(\bm{\tau})$. Let $N_c$ be the number of constraints. Then, $N_c$ should not exceed $N$. For convenience, $N_c$ constrained grids are labeled by $i$. We first define $N_c\times N_c$-matrix $\hat{A}$, $N_c\times(N-N_c)$-matrix $\hat{B}$, and $N_c$-dimensional vector $\bm{D}$ as
\begin{align}
    A_{i,l} &= g_l(\bm{\tau}_i),\quad (l\leq N_c) \\
    B_{i,l-N_c} &= g_l(\bm{\tau}_i),\quad (l>N_c) \\
    D_i &= D(\bm{\tau}_i).
\end{align}
Using these entities, $N_c$-dimensional vector $\bm{C}^{(0)}$ and $(N-N_c)\times N_c$-matrix $\bm{C}^{(1)}$ are defined as
\begin{align}
    \bm{C}^{(0)} &= \hat{A}^{-1}\bm{D}, \\
    \hat{C}^{(1)} &= {}^t(\hat{A}^{-1}\hat{B}).
\end{align}
Now, new functions $\tilde{g}_l(\bm{\tau})$ for $l>N_c$ are introduced as
\begin{equation}
    \tilde{g}_l(\bm{\tau}) = g_l(\bm{\tau}) -\sum_{l'=1}^{N_c} C^{(1)}_{l-N_c,l'}g_{l'}(\bm{\tau}),
\end{equation}
and the constrained function $Q(\bm{\tau})$ is written as
\begin{equation}
    Q(\bm{\tau}) = \sum_{l=1}^{N_c}C^{(0)}_lg_l(\bm{\tau}) + \sum_{l>N_c}^Nc_l\tilde{g}_l(\bm{\tau}).
\end{equation}
Note that we have $\tilde{g}_l(\bm{\tau}_i)=0$, which is confirmed as
\begin{equation}
\begin{split}
    \tilde{g}_l(\bm{\tau}_i) &= B_{i,l-N_c}-\sum_{l'=1}^{N_c}(\hat{A}^{-1}\hat{B})_{l',l-N_c}A_{i,l'}\\
    &= B_{i,l-N_c} -(\hat{A}\hat{A}^{-1}\hat{B})_{i,l-N_c} = 0,
\end{split}
\end{equation}
and we have $Q(\bm{\tau}_i)=D(\bm{\tau}_i)$, which is confirmed as
\begin{equation}
    \begin{split}
        Q(\bm{\tau}_i) &= \sum_{l=1}^{N_c}(\hat{A}^{-1}\bm{D})_lA_{i,l} \\
        &= (\hat{A}\hat{A}^{-1}\bm{D})_i = D(\bm{\tau}_i).
    \end{split}
\end{equation}

To minimize $\Xi$, we take derivative of $\Xi$ with respect to $c_l$, leading to
\begin{equation}
    \frac{\partial \Xi}{\partial c_l} 
    = 2\sum_{l'}M_{l,l'}c_{l'}-2X_l,
\end{equation}
where 
\begin{align}
    M_{l,l'} &= \sum_{\bm{\tau}\in\text{grid}}\tilde{g}_l(\bm{\tau})\tilde{g}_{l'}(\bm{\tau}),\\
    X_l &= \sum_{\bm{\tau}\in\text{grid}}\tilde{g}_l(\bm{\tau})(D(\bm{\tau})-\sum_{l'}C^{(0)}_{l'}g_{l'}(\bm{\tau})).
\end{align}
Then, the optimized $c_l$ can be obtained from 
\begin{equation}
    \bm{c} = \hat{M}^{-1}\bm{X}.
\end{equation}

\section{Interlayer distance}
\label{app:dz}
To obtain $E_{\pm}(\bm{k}_M,\bm{\tau})$ within DFT, we need to estimate the interlayer distance $d_z$ for each $\bm{\tau}$. We first derive $d_z$ dependence of the total energy within DFT, and choose $d_z$ that realizes minimum total energy. While scanning over $d_z$, the intralayer coordinates are fixed. Note that if in-plane relaxation is also allowed, $\bm{\tau}$ can flow into the one with lower total energy, which prevents us from getting $\bm{\tau}$ dependence. In practice, $d_z$-scan is done over finite set of $d_z$ values. Then, for each $\bm{\tau}$, the obtained data are used to fit an approximated expression 
\begin{equation}
    E_{\text{tot}}(d_z) = \alpha e^{-\beta(d_z-d_0)}-\gamma\Bigl(\frac{d_0}{d_z}\Bigr)^\delta.
\end{equation}
Here, $d_0$ is chosen to be 4\AA, and $\{\alpha,\beta,\gamma,\delta\}$ are the adjustable parameters in the fitting. The fitting is done with FindFit function in Mathematica.

\begin{figure}
    \centering
    \includegraphics[]{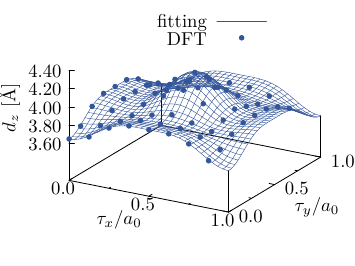}
    \caption{Interlayer distance obtained by the lower harmonics expansion (lines) and DFT (points).}
    \label{fig:map_dz}
\end{figure}

The $\bm{\tau}$ dependence is approximated by 
\begin{equation}
\begin{split}
    d_z(\bm{\tau}) = c_0 &+ c_1(\cos\tilde{\tau}_x+\cos\tilde{\tau}_y) \\
    &+c_2(\cos(\tilde{\tau}_x+\tilde{\tau}_y)-\cos(\tilde{\tau}_x+\tilde{\tau}_y))\\
    &+c_3(\cos 2\tilde{\tau}_x+\cos 2\tilde{\tau}_y)\\    &+c_4(\cos(2\tilde{\tau}_x+\tilde{\tau}_y)+\cos(\tilde{\tau}_x+2\tilde{\tau}_y)\\
    &\quad\quad +\cos(2\tilde{\tau}_x-\tilde{\tau}_y)+\cos(\tilde{\tau}_x-2\tilde{\tau}_y)).
\end{split}
\end{equation}
The coefficients are fixed by the constrained least square algorithm above. Here, DFT calculations are done on $8\times 8$ regular grids in a unit cell, and the constraint is such that the fitting faithfully reproduce the DFT results at $\bm{\tau}=(0,0)$, $(a_0/2,0)$, and $(a_0/2,a_0/2)$, which correspond to the unit cell center, the center of the unit cell edge, and the unit cell corner, respectively. Using the obtained coefficients,
\begin{gather}
    c_0 = 3.896,\, c_1 = -0.093,\, c_2=0.007 \\
    c_3 = 0.021,\, c_4 = -0.028,
\end{gather}
which are in the unit of \AA, the fitting and the DFT results compare as in Fig.~\ref{fig:map_dz}, showing satisfactory match between them. 

\section{Emergent symmetry $\mathcal{L}$}
\label{app:symmetry}
In the convenient basis described near equation \eqref{eq:gauge_choice}, the Hamiltonian is
\be
H'(\bm{q})=\begin{pmatrix} H_-'(\bm{q})+V_0+V_3  & e^{\mathrm{i}\delta \bm{k}\cdot \bm{r}} V_1(\bm{r}) \\
e^{-\mathrm{i}\delta \bm{k}\cdot \bm{r}} V_1(\bm{r}) & H_+'(\bm{q})+V_0-V_3 \end{pmatrix}. 
\ee
Under $C_{4z}$ rotation, different terms in the Hamiltonian transforms as %we have $\bm{r}\rightarrow R_{\pi/2}\bm{r},\quad (-\mathrm{i}\bm{\nabla})\rightarrow R_4 (-\mathrm{i}\bm{\nabla})$ and 
\be
\begin{split}
& C_{4z} H'_+(\bm{q}) C_{4z}^{-1} \\
= & \frac{\hbar^2}{2m_*} \left[ 
-i\bm{\nabla}+ R_{-\pi/2}\bm{q}+ R_{-\pi/2}\bm{q}_1/2
\right]^2 \\
= & \frac{\hbar^2}{2m_*} \left[ 
-i\bm{\nabla}+ R_{-\pi/2}\bm{q} -\bm{q}_2/2
\right]^2 \\
= & e^{\mathrm{i}\bm{q}_2\cdot \bm{r}} H_-' (R_{-\pi/2}\bm{q}) e^{-\mathrm{i}\bm{q}_2\cdot \bm{r}}. 
\end{split}
\ee
Similarly, 
\be
\begin{split}
& C_{4z} H'_-(\bm{q}) C_{4z}^{-1} \\
= & \frac{\hbar^2}{2m_*} \left[ 
-i\bm{\nabla}+ R_{-\pi/2}\bm{q}+ R_{-\pi/2}\bm{q}_2/2
\right]^2 \\
= & \frac{\hbar^2}{2m_*} \left[ 
-i\bm{\nabla}+ R_{-\pi/2}\bm{q} +\bm{q}_1/2
\right]^2 = H'_+ (R_{-\pi/2} \bm{q}).
\end{split}
\ee
As for the moiré potentials,
\be
V_{0,1} (R_{\pi/2} \bm{r}) =V_{0,1} (\bm{r}), \quad V_3 (R_{\pi/2} \bm{r}) = -V_3 (\bm{r}). 
\ee
Putting everything together, the whole Hamiltonian transforms as
\be
\begin{split}
& C_{4z}\tilde{H}(\bm{q}) C_{4z}^{-1} =V_0 \sigma_0 -V_3 \sigma_3\\
& \quad \quad+  \begin{pmatrix}
H_+'(R_{-\pi/2}\bm{q}) & e^{\mathrm{i}\delta\bm{k}\cdot R_{\pi/2} \bm{r}} V_1 \\
e^{-\mathrm{i}\delta \bm{k}\cdot R_{\pi/2}\bm{r}} V_1 & e^{\mathrm{i}\bm{q}_2\cdot \bm{r}} H_-'(R_{-\pi/2} \bm{q} )e^{-\mathrm{i}\bm{q}_2\cdot \bm{r}} )
\end{pmatrix}
\end{split}
\ee
The second line can be further simplified to
\be
\begin{split}
& \begin{pmatrix}
H_+'(R_{-\pi/2}\bm{q}) & e^{\mathrm{i} R_{-\pi/2} \delta\bm{k}\cdot \bm{r}} V_1 \\
e^{-\mathrm{i} R_{-\pi/2}\delta\bm{k}\cdot \bm{r}} V_1 & e^{\mathrm{i}\bm{q}_2\cdot \bm{r}} H_-'(R_{-\pi/2} \bm{q}) e^{-\mathrm{i}\bm{q}_2\cdot \bm{r}} )
\end{pmatrix}\\
= & \begin{pmatrix}
H_+'(R_{-\pi/2}\bm{q}) & e^{-\mathrm{i}  \delta\bm{k}\cdot \bm{r}} e^{-\mathrm{i}  \bm{q}_2 \cdot \bm{r}} V_1 \\
e^{+\mathrm{i} \delta\bm{k}\cdot \bm{r}}e^{\mathrm{i}  \bm{q}_2 \cdot \bm{r}} V_1 & e^{\mathrm{i}\bm{q}_2\cdot \bm{r}} H_-'(R_{-\pi/2} \bm{q}) e^{-\mathrm{i}\bm{q}_2\cdot \bm{r}} )
\end{pmatrix}
\end{split}
\ee
where we have used
\be
R_{-\pi/2} \delta \bm{k}= (-\bm{q}_1-\bm{q}_2)/2 = -\delta \bm{k}-\bm{q}_2. 
\ee
Further, applying $e^{-\frac{\mathrm{i}}{2}\bm{q}_2\cdot \bm{r} (\sigma_3+\sigma_0)}\sigma_1$, we obtain
\be
\begin{split}
\mathcal{L}\tilde{H}(\bm{q}) \mathcal{L}^{-1} = 
& e^{-\frac{\mathrm{i}}{2}\bm{q}_2\cdot \bm{r} (\sigma_3+\sigma_0)}\sigma_1 C_{4z} \tilde{H}(\bm{q}) C_{4z}^{-1} \sigma_1 e^{\frac{\mathrm{i}}{2}\bm{q}_2\cdot \bm{r} (\sigma_3+\sigma_0)}\\
= & \tilde{H} (R_{-\pi/2} \bm{q}).
\end{split}
\ee

\bibliography{draft_ref.bib}

\begin{thebibliography}{51}%
\makeatletter
\providecommand \@ifxundefined [1]{%
 \@ifx{#1\undefined}
}%
\providecommand \@ifnum [1]{%
 \ifnum #1\expandafter \@firstoftwo
 \else \expandafter \@secondoftwo
 \fi
}%
\providecommand \@ifx [1]{%
 \ifx #1\expandafter \@firstoftwo
 \else \expandafter \@secondoftwo
 \fi
}%
\providecommand \natexlab [1]{#1}%
\providecommand \enquote  [1]{``#1''}%
\providecommand \bibnamefont  [1]{#1}%
\providecommand \bibfnamefont [1]{#1}%
\providecommand \citenamefont [1]{#1}%
\providecommand \href@noop [0]{\@secondoftwo}%
\providecommand \href [0]{\begingroup \@sanitize@url \@href}%
\providecommand \@href[1]{\@@startlink{#1}\@@href}%
\providecommand \@@href[1]{\endgroup#1\@@endlink}%
\providecommand \@sanitize@url [0]{\catcode `\\12\catcode `\$12\catcode `\&12\catcode `\#12\catcode `\^12\catcode `\_12\catcode `\%12\relax}%
\providecommand \@@startlink[1]{}%
\providecommand \@@endlink[0]{}%
\providecommand \url  [0]{\begingroup\@sanitize@url \@url }%
\providecommand \@url [1]{\endgroup\@href {#1}{\urlprefix }}%
\providecommand \urlprefix  [0]{URL }%
\providecommand \Eprint [0]{\href }%
\providecommand \doibase [0]{https://doi.org/}%
\providecommand \selectlanguage [0]{\@gobble}%
\providecommand \bibinfo  [0]{\@secondoftwo}%
\providecommand \bibfield  [0]{\@secondoftwo}%
\providecommand \translation [1]{[#1]}%
\providecommand \BibitemOpen [0]{}%
\providecommand \bibitemStop [0]{}%
\providecommand \bibitemNoStop [0]{.\EOS\space}%
\providecommand \EOS [0]{\spacefactor3000\relax}%
\providecommand \BibitemShut  [1]{\csname bibitem#1\endcsname}%
\let\auto@bib@innerbib\@empty
%</preamble>
\bibitem [{\citenamefont {Keimer}\ \emph {et~al.}(2015)\citenamefont {Keimer}, \citenamefont {Kivelson}, \citenamefont {Norman}, \citenamefont {Uchida},\ and\ \citenamefont {Zaanen}}]{keimer2015quantum}%
  \BibitemOpen
  \bibfield  {author} {\bibinfo {author} {\bibfnamefont {B.}~\bibnamefont {Keimer}}, \bibinfo {author} {\bibfnamefont {S.~A.}\ \bibnamefont {Kivelson}}, \bibinfo {author} {\bibfnamefont {M.~R.}\ \bibnamefont {Norman}}, \bibinfo {author} {\bibfnamefont {S.}~\bibnamefont {Uchida}},\ and\ \bibinfo {author} {\bibfnamefont {J.}~\bibnamefont {Zaanen}},\ }\bibfield  {title} {\bibinfo {title} {From quantum matter to high-temperature superconductivity in copper oxides},\ }\href@noop {} {\bibfield  {journal} {\bibinfo  {journal} {Nature}\ }\textbf {\bibinfo {volume} {518}},\ \bibinfo {pages} {179} (\bibinfo {year} {2015})}\BibitemShut {NoStop}%
\bibitem [{\citenamefont {Kennes}\ \emph {et~al.}(2021)\citenamefont {Kennes}, \citenamefont {Claassen}, \citenamefont {Xian}, \citenamefont {Georges}, \citenamefont {Millis}, \citenamefont {Hone}, \citenamefont {Dean}, \citenamefont {Basov}, \citenamefont {Pasupathy},\ and\ \citenamefont {Rubio}}]{kennes2021moire}%
  \BibitemOpen
  \bibfield  {author} {\bibinfo {author} {\bibfnamefont {D.~M.}\ \bibnamefont {Kennes}}, \bibinfo {author} {\bibfnamefont {M.}~\bibnamefont {Claassen}}, \bibinfo {author} {\bibfnamefont {L.}~\bibnamefont {Xian}}, \bibinfo {author} {\bibfnamefont {A.}~\bibnamefont {Georges}}, \bibinfo {author} {\bibfnamefont {A.~J.}\ \bibnamefont {Millis}}, \bibinfo {author} {\bibfnamefont {J.}~\bibnamefont {Hone}}, \bibinfo {author} {\bibfnamefont {C.~R.}\ \bibnamefont {Dean}}, \bibinfo {author} {\bibfnamefont {D.}~\bibnamefont {Basov}}, \bibinfo {author} {\bibfnamefont {A.~N.}\ \bibnamefont {Pasupathy}},\ and\ \bibinfo {author} {\bibfnamefont {A.}~\bibnamefont {Rubio}},\ }\bibfield  {title} {\bibinfo {title} {Moir{\'e} heterostructures as a condensed-matter quantum simulator},\ }\href@noop {} {\bibfield  {journal} {\bibinfo  {journal} {Nature Physics}\ }\textbf {\bibinfo {volume} {17}},\ \bibinfo {pages} {155} (\bibinfo {year} {2021})}\BibitemShut {NoStop}%
\bibitem [{\citenamefont {Wu}\ \emph {et~al.}(2018)\citenamefont {Wu}, \citenamefont {Lovorn}, \citenamefont {Tutuc},\ and\ \citenamefont {MacDonald}}]{PhysRevLett.121.026402}%
  \BibitemOpen
  \bibfield  {author} {\bibinfo {author} {\bibfnamefont {F.}~\bibnamefont {Wu}}, \bibinfo {author} {\bibfnamefont {T.}~\bibnamefont {Lovorn}}, \bibinfo {author} {\bibfnamefont {E.}~\bibnamefont {Tutuc}},\ and\ \bibinfo {author} {\bibfnamefont {A.~H.}\ \bibnamefont {MacDonald}},\ }\bibfield  {title} {\bibinfo {title} {Hubbard model physics in transition metal dichalcogenide moir\'e bands},\ }\href {https://doi.org/10.1103/PhysRevLett.121.026402} {\bibfield  {journal} {\bibinfo  {journal} {Phys. Rev. Lett.}\ }\textbf {\bibinfo {volume} {121}},\ \bibinfo {pages} {026402} (\bibinfo {year} {2018})}\BibitemShut {NoStop}%
\bibitem [{\citenamefont {Tang}\ \emph {et~al.}(2020)\citenamefont {Tang}, \citenamefont {Li}, \citenamefont {Li}, \citenamefont {Xu}, \citenamefont {Liu}, \citenamefont {Barmak}, \citenamefont {Watanabe}, \citenamefont {Taniguchi}, \citenamefont {MacDonald}, \citenamefont {Shan} \emph {et~al.}}]{tang2020simulation}%
  \BibitemOpen
  \bibfield  {author} {\bibinfo {author} {\bibfnamefont {Y.}~\bibnamefont {Tang}}, \bibinfo {author} {\bibfnamefont {L.}~\bibnamefont {Li}}, \bibinfo {author} {\bibfnamefont {T.}~\bibnamefont {Li}}, \bibinfo {author} {\bibfnamefont {Y.}~\bibnamefont {Xu}}, \bibinfo {author} {\bibfnamefont {S.}~\bibnamefont {Liu}}, \bibinfo {author} {\bibfnamefont {K.}~\bibnamefont {Barmak}}, \bibinfo {author} {\bibfnamefont {K.}~\bibnamefont {Watanabe}}, \bibinfo {author} {\bibfnamefont {T.}~\bibnamefont {Taniguchi}}, \bibinfo {author} {\bibfnamefont {A.~H.}\ \bibnamefont {MacDonald}}, \bibinfo {author} {\bibfnamefont {J.}~\bibnamefont {Shan}}, \emph {et~al.},\ }\bibfield  {title} {\bibinfo {title} {Simulation of hubbard model physics in wse2/ws2 moir{\'e} superlattices},\ }\href@noop {} {\bibfield  {journal} {\bibinfo  {journal} {Nature}\ }\textbf {\bibinfo {volume} {579}},\ \bibinfo {pages} {353} (\bibinfo {year} {2020})}\BibitemShut {NoStop}%
\bibitem [{\citenamefont {Proust}\ and\ \citenamefont {Taillefer}(2019)}]{review_cuprates}%
  \BibitemOpen
  \bibfield  {author} {\bibinfo {author} {\bibfnamefont {C.}~\bibnamefont {Proust}}\ and\ \bibinfo {author} {\bibfnamefont {L.}~\bibnamefont {Taillefer}},\ }\bibfield  {title} {\bibinfo {title} {The remarkable underlying ground states of cuprate superconductors},\ }\href {https://doi.org/https://doi.org/10.1146/annurev-conmatphys-031218-013210} {\bibfield  {journal} {\bibinfo  {journal} {Annual Review of Condensed Matter Physics}\ }\textbf {\bibinfo {volume} {10}},\ \bibinfo {pages} {409} (\bibinfo {year} {2019})}\BibitemShut {NoStop}%
\bibitem [{\citenamefont {Xu}\ \emph {et~al.}(2024)\citenamefont {Xu}, \citenamefont {Tancogne-Dejean}, \citenamefont {Boström}, \citenamefont {Kennes}, \citenamefont {Claassen}, \citenamefont {Rubio},\ and\ \citenamefont {Xian}}]{xu2024engineering2dsquarelattice}%
  \BibitemOpen
  \bibfield  {author} {\bibinfo {author} {\bibfnamefont {Q.}~\bibnamefont {Xu}}, \bibinfo {author} {\bibfnamefont {N.}~\bibnamefont {Tancogne-Dejean}}, \bibinfo {author} {\bibfnamefont {E.~V.}\ \bibnamefont {Boström}}, \bibinfo {author} {\bibfnamefont {D.~M.}\ \bibnamefont {Kennes}}, \bibinfo {author} {\bibfnamefont {M.}~\bibnamefont {Claassen}}, \bibinfo {author} {\bibfnamefont {A.}~\bibnamefont {Rubio}},\ and\ \bibinfo {author} {\bibfnamefont {L.}~\bibnamefont {Xian}},\ }\href {https://arxiv.org/abs/2406.05626} {\bibinfo {title} {Engineering 2d square lattice hubbard models in 90$^\circ$ twisted ge/snx (x=s, se) moir\'e supperlattices}} (\bibinfo {year} {2024}),\ \Eprint {https://arxiv.org/abs/2406.05626} {arXiv:2406.05626 [cond-mat.str-el]} \BibitemShut {NoStop}%
\bibitem [{\citenamefont {Kariyado}\ and\ \citenamefont {Vishwanath}(2019)}]{PhysRevResearch.1.033076}%
  \BibitemOpen
  \bibfield  {author} {\bibinfo {author} {\bibfnamefont {T.}~\bibnamefont {Kariyado}}\ and\ \bibinfo {author} {\bibfnamefont {A.}~\bibnamefont {Vishwanath}},\ }\bibfield  {title} {\bibinfo {title} {Flat band in twisted bilayer {Bravais} lattices},\ }\href {https://doi.org/10.1103/PhysRevResearch.1.033076} {\bibfield  {journal} {\bibinfo  {journal} {Phys. Rev. Research}\ }\textbf {\bibinfo {volume} {1}},\ \bibinfo {pages} {033076} (\bibinfo {year} {2019})}\BibitemShut {NoStop}%
\bibitem [{\citenamefont {Eugenio}\ \emph {et~al.}(2024)\citenamefont {Eugenio}, \citenamefont {Luo}, \citenamefont {Vishwanath},\ and\ \citenamefont {Volkov}}]{Pavel}%
  \BibitemOpen
  \bibfield  {author} {\bibinfo {author} {\bibfnamefont {P.~M.}\ \bibnamefont {Eugenio}}, \bibinfo {author} {\bibfnamefont {Z.-X.}\ \bibnamefont {Luo}}, \bibinfo {author} {\bibfnamefont {A.}~\bibnamefont {Vishwanath}},\ and\ \bibinfo {author} {\bibfnamefont {P.~A.}\ \bibnamefont {Volkov}},\ }\href {https://arxiv.org/abs/2406.02448} {\bibinfo {title} {Tunable {$t-t'-U$} hubbard models in twisted square homobilayers}} (\bibinfo {year} {2024}),\ \Eprint {https://arxiv.org/abs/2406.02448} {arXiv:2406.02448 [cond-mat.str-el]} \BibitemShut {NoStop}%
\bibitem [{\citenamefont {Ji}\ \emph {et~al.}(2019)\citenamefont {Ji}, \citenamefont {Cai}, \citenamefont {Paudel}, \citenamefont {Sun}, \citenamefont {Zhang}, \citenamefont {Han}, \citenamefont {Wei}, \citenamefont {Zang}, \citenamefont {Gu}, \citenamefont {Zhang} \emph {et~al.}}]{ji2019freestanding}%
  \BibitemOpen
  \bibfield  {author} {\bibinfo {author} {\bibfnamefont {D.}~\bibnamefont {Ji}}, \bibinfo {author} {\bibfnamefont {S.}~\bibnamefont {Cai}}, \bibinfo {author} {\bibfnamefont {T.~R.}\ \bibnamefont {Paudel}}, \bibinfo {author} {\bibfnamefont {H.}~\bibnamefont {Sun}}, \bibinfo {author} {\bibfnamefont {C.}~\bibnamefont {Zhang}}, \bibinfo {author} {\bibfnamefont {L.}~\bibnamefont {Han}}, \bibinfo {author} {\bibfnamefont {Y.}~\bibnamefont {Wei}}, \bibinfo {author} {\bibfnamefont {Y.}~\bibnamefont {Zang}}, \bibinfo {author} {\bibfnamefont {M.}~\bibnamefont {Gu}}, \bibinfo {author} {\bibfnamefont {Y.}~\bibnamefont {Zhang}}, \emph {et~al.},\ }\bibfield  {title} {\bibinfo {title} {Freestanding crystalline oxide perovskites down to the monolayer limit},\ }\href@noop {} {\bibfield  {journal} {\bibinfo  {journal} {Nature}\ }\textbf {\bibinfo {volume} {570}},\ \bibinfo {pages} {87} (\bibinfo {year} {2019})}\BibitemShut {NoStop}%
\bibitem [{\citenamefont {Shigekawa}\ \emph {et~al.}(2019)\citenamefont {Shigekawa}, \citenamefont {Nakayama}, \citenamefont {Kuno}, \citenamefont {Phan}, \citenamefont {Owada}, \citenamefont {Sugawara}, \citenamefont {Takahashi},\ and\ \citenamefont {Sato}}]{shigekawa2019dichotomy}%
  \BibitemOpen
  \bibfield  {author} {\bibinfo {author} {\bibfnamefont {K.}~\bibnamefont {Shigekawa}}, \bibinfo {author} {\bibfnamefont {K.}~\bibnamefont {Nakayama}}, \bibinfo {author} {\bibfnamefont {M.}~\bibnamefont {Kuno}}, \bibinfo {author} {\bibfnamefont {G.~N.}\ \bibnamefont {Phan}}, \bibinfo {author} {\bibfnamefont {K.}~\bibnamefont {Owada}}, \bibinfo {author} {\bibfnamefont {K.}~\bibnamefont {Sugawara}}, \bibinfo {author} {\bibfnamefont {T.}~\bibnamefont {Takahashi}},\ and\ \bibinfo {author} {\bibfnamefont {T.}~\bibnamefont {Sato}},\ }\bibfield  {title} {\bibinfo {title} {Dichotomy of superconductivity between monolayer fes and fese},\ }\href@noop {} {\bibfield  {journal} {\bibinfo  {journal} {Proceedings of the National Academy of Sciences}\ }\textbf {\bibinfo {volume} {116}},\ \bibinfo {pages} {24470} (\bibinfo {year} {2019})}\BibitemShut {NoStop}%
\bibitem [{\citenamefont {Yu}\ \emph {et~al.}(2019)\citenamefont {Yu}, \citenamefont {Ma}, \citenamefont {Cai}, \citenamefont {Zhong}, \citenamefont {Ye}, \citenamefont {Shen}, \citenamefont {Gu}, \citenamefont {Chen},\ and\ \citenamefont {Zhang}}]{yu2019high}%
  \BibitemOpen
  \bibfield  {author} {\bibinfo {author} {\bibfnamefont {Y.}~\bibnamefont {Yu}}, \bibinfo {author} {\bibfnamefont {L.}~\bibnamefont {Ma}}, \bibinfo {author} {\bibfnamefont {P.}~\bibnamefont {Cai}}, \bibinfo {author} {\bibfnamefont {R.}~\bibnamefont {Zhong}}, \bibinfo {author} {\bibfnamefont {C.}~\bibnamefont {Ye}}, \bibinfo {author} {\bibfnamefont {J.}~\bibnamefont {Shen}}, \bibinfo {author} {\bibfnamefont {G.~D.}\ \bibnamefont {Gu}}, \bibinfo {author} {\bibfnamefont {X.~H.}\ \bibnamefont {Chen}},\ and\ \bibinfo {author} {\bibfnamefont {Y.}~\bibnamefont {Zhang}},\ }\bibfield  {title} {\bibinfo {title} {High-temperature superconductivity in monolayer bi2sr2cacu2o8+ $\delta$},\ }\href@noop {} {\bibfield  {journal} {\bibinfo  {journal} {Nature}\ }\textbf {\bibinfo {volume} {575}},\ \bibinfo {pages} {156} (\bibinfo {year} {2019})}\BibitemShut {NoStop}%
\bibitem [{\citenamefont {Wang}\ \emph {et~al.}(2012)\citenamefont {Wang}, \citenamefont {Li}, \citenamefont {Zhang}, \citenamefont {Zhang}, \citenamefont {Zhang}, \citenamefont {Li}, \citenamefont {Ding}, \citenamefont {Ou}, \citenamefont {Deng}, \citenamefont {Chang} \emph {et~al.}}]{Wang_2012}%
  \BibitemOpen
  \bibfield  {author} {\bibinfo {author} {\bibfnamefont {Q.-Y.}\ \bibnamefont {Wang}}, \bibinfo {author} {\bibfnamefont {Z.}~\bibnamefont {Li}}, \bibinfo {author} {\bibfnamefont {W.-H.}\ \bibnamefont {Zhang}}, \bibinfo {author} {\bibfnamefont {Z.-C.}\ \bibnamefont {Zhang}}, \bibinfo {author} {\bibfnamefont {J.-S.}\ \bibnamefont {Zhang}}, \bibinfo {author} {\bibfnamefont {W.}~\bibnamefont {Li}}, \bibinfo {author} {\bibfnamefont {H.}~\bibnamefont {Ding}}, \bibinfo {author} {\bibfnamefont {Y.-B.}\ \bibnamefont {Ou}}, \bibinfo {author} {\bibfnamefont {P.}~\bibnamefont {Deng}}, \bibinfo {author} {\bibfnamefont {K.}~\bibnamefont {Chang}}, \emph {et~al.},\ }\bibfield  {title} {\bibinfo {title} {Interface-induced high-temperature superconductivity in single unit-cell fese films on {SrTiO3}},\ }\href@noop {} {\bibfield  {journal} {\bibinfo  {journal} {Chinese Physics Letters}\ }\textbf {\bibinfo {volume} {29}},\ \bibinfo {pages} {037402} (\bibinfo {year} {2012})}\BibitemShut {NoStop}%
\bibitem [{\citenamefont {Mounet}\ \emph {et~al.}(2018)\citenamefont {Mounet}, \citenamefont {Gibertini}, \citenamefont {Schwaller}, \citenamefont {Campi}, \citenamefont {Merkys}, \citenamefont {Marrazzo}, \citenamefont {Sohier}, \citenamefont {Castelli}, \citenamefont {Cepellotti}, \citenamefont {Pizzi} \emph {et~al.}}]{mounet2018two}%
  \BibitemOpen
  \bibfield  {author} {\bibinfo {author} {\bibfnamefont {N.}~\bibnamefont {Mounet}}, \bibinfo {author} {\bibfnamefont {M.}~\bibnamefont {Gibertini}}, \bibinfo {author} {\bibfnamefont {P.}~\bibnamefont {Schwaller}}, \bibinfo {author} {\bibfnamefont {D.}~\bibnamefont {Campi}}, \bibinfo {author} {\bibfnamefont {A.}~\bibnamefont {Merkys}}, \bibinfo {author} {\bibfnamefont {A.}~\bibnamefont {Marrazzo}}, \bibinfo {author} {\bibfnamefont {T.}~\bibnamefont {Sohier}}, \bibinfo {author} {\bibfnamefont {I.~E.}\ \bibnamefont {Castelli}}, \bibinfo {author} {\bibfnamefont {A.}~\bibnamefont {Cepellotti}}, \bibinfo {author} {\bibfnamefont {G.}~\bibnamefont {Pizzi}}, \emph {et~al.},\ }\bibfield  {title} {\bibinfo {title} {Two-dimensional materials from high-throughput computational exfoliation of experimentally known compounds},\ }\href@noop {} {\bibfield  {journal} {\bibinfo  {journal} {Nature nanotechnology}\ }\textbf {\bibinfo {volume} {13}},\ \bibinfo {pages} {246} (\bibinfo {year} {2018})}\BibitemShut {NoStop}%
\bibitem [{\citenamefont {Ram}\ and\ \citenamefont {Mizuseki}(2020)}]{RAM2020827}%
  \BibitemOpen
  \bibfield  {author} {\bibinfo {author} {\bibfnamefont {B.}~\bibnamefont {Ram}}\ and\ \bibinfo {author} {\bibfnamefont {H.}~\bibnamefont {Mizuseki}},\ }\bibfield  {title} {\bibinfo {title} {{$C_{568}$}: A new two-dimensional ${sp^2-sp^3}$ hybridized allotrope of carbon},\ }\href {https://doi.org/https://doi.org/10.1016/j.carbon.2019.11.062} {\bibfield  {journal} {\bibinfo  {journal} {Carbon}\ }\textbf {\bibinfo {volume} {158}},\ \bibinfo {pages} {827} (\bibinfo {year} {2020})}\BibitemShut {NoStop}%
\bibitem [{\citenamefont {Gao}\ \emph {et~al.}(2020)\citenamefont {Gao}, \citenamefont {Sahin},\ and\ \citenamefont {Kang}}]{20050034}%
  \BibitemOpen
  \bibfield  {author} {\bibinfo {author} {\bibfnamefont {Q.}~\bibnamefont {Gao}}, \bibinfo {author} {\bibfnamefont {H.}~\bibnamefont {Sahin}},\ and\ \bibinfo {author} {\bibfnamefont {J.}~\bibnamefont {Kang}},\ }\href {https://doi.org/10.1088/1674-4926/41/8/082005} {\bibinfo {title} {Strain tunable band structure of a new 2d carbon allotrope {$C_{568}$}}} (\bibinfo {year} {2020})\BibitemShut {NoStop}%
\bibitem [{\citenamefont {Lu}\ \emph {et~al.}(2022)\citenamefont {Lu}, \citenamefont {Wang}, \citenamefont {Li}, \citenamefont {Dai}, \citenamefont {Yin},\ and\ \citenamefont {Ding}}]{doi:10.1080/1536383X.2021.1944119}%
  \BibitemOpen
  \bibfield  {author} {\bibinfo {author} {\bibfnamefont {K.}~\bibnamefont {Lu}}, \bibinfo {author} {\bibfnamefont {T.}~\bibnamefont {Wang}}, \bibinfo {author} {\bibfnamefont {X.}~\bibnamefont {Li}}, \bibinfo {author} {\bibfnamefont {L.}~\bibnamefont {Dai}}, \bibinfo {author} {\bibfnamefont {J.}~\bibnamefont {Yin}},\ and\ \bibinfo {author} {\bibfnamefont {Y.}~\bibnamefont {Ding}},\ }\bibfield  {title} {\bibinfo {title} {A new 2d carbon allotrope {$C_{568}$} as a high-capacity electrode material for lithium-ion batteries},\ }\href {https://doi.org/10.1080/1536383X.2021.1944119} {\bibfield  {journal} {\bibinfo  {journal} {Fullerenes, Nanotubes and Carbon Nanostructures}\ }\textbf {\bibinfo {volume} {30}},\ \bibinfo {pages} {385} (\bibinfo {year} {2022})},\ \Eprint {https://arxiv.org/abs/https://doi.org/10.1080/1536383X.2021.1944119} {https://doi.org/10.1080/1536383X.2021.1944119} \BibitemShut {NoStop}%
\bibitem [{\citenamefont {Zhao}\ \emph {et~al.}(2023{\natexlab{a}})\citenamefont {Zhao}, \citenamefont {Li}, \citenamefont {Zhang}, \citenamefont {Eglitis}, \citenamefont {Wang},\ and\ \citenamefont {Jia}}]{ZHAO2023154895}%
  \BibitemOpen
  \bibfield  {author} {\bibinfo {author} {\bibfnamefont {W.-H.}\ \bibnamefont {Zhao}}, \bibinfo {author} {\bibfnamefont {F.-Y.}\ \bibnamefont {Li}}, \bibinfo {author} {\bibfnamefont {H.-X.}\ \bibnamefont {Zhang}}, \bibinfo {author} {\bibfnamefont {R.~I.}\ \bibnamefont {Eglitis}}, \bibinfo {author} {\bibfnamefont {J.}~\bibnamefont {Wang}},\ and\ \bibinfo {author} {\bibfnamefont {R.}~\bibnamefont {Jia}},\ }\bibfield  {title} {\bibinfo {title} {Doping at sp3-site in me-graphene (c568) for new anodes in rechargeable li-ion battery},\ }\href {https://doi.org/https://doi.org/10.1016/j.apsusc.2022.154895} {\bibfield  {journal} {\bibinfo  {journal} {Applied Surface Science}\ }\textbf {\bibinfo {volume} {607}},\ \bibinfo {pages} {154895} (\bibinfo {year} {2023}{\natexlab{a}})}\BibitemShut {NoStop}%
\bibitem [{\citenamefont {Jiang}\ and\ \citenamefont {Devereaux}(2019)}]{doi:10.1126/science.aal5304}%
  \BibitemOpen
  \bibfield  {author} {\bibinfo {author} {\bibfnamefont {H.-C.}\ \bibnamefont {Jiang}}\ and\ \bibinfo {author} {\bibfnamefont {T.~P.}\ \bibnamefont {Devereaux}},\ }\bibfield  {title} {\bibinfo {title} {Superconductivity in the doped hubbard model and its interplay with next-nearest hopping $t'$},\ }\href {https://doi.org/10.1126/science.aal5304} {\bibfield  {journal} {\bibinfo  {journal} {Science}\ }\textbf {\bibinfo {volume} {365}},\ \bibinfo {pages} {1424} (\bibinfo {year} {2019})},\ \Eprint {https://arxiv.org/abs/https://www.science.org/doi/pdf/10.1126/science.aal5304} {https://www.science.org/doi/pdf/10.1126/science.aal5304} \BibitemShut {NoStop}%
\bibitem [{\citenamefont {Chung}\ \emph {et~al.}(2020)\citenamefont {Chung}, \citenamefont {Qin}, \citenamefont {Zhang}, \citenamefont {Schollw\"ock},\ and\ \citenamefont {White}}]{PhysRevB.102.041106}%
  \BibitemOpen
  \bibfield  {author} {\bibinfo {author} {\bibfnamefont {C.-M.}\ \bibnamefont {Chung}}, \bibinfo {author} {\bibfnamefont {M.}~\bibnamefont {Qin}}, \bibinfo {author} {\bibfnamefont {S.}~\bibnamefont {Zhang}}, \bibinfo {author} {\bibfnamefont {U.}~\bibnamefont {Schollw\"ock}},\ and\ \bibinfo {author} {\bibfnamefont {S.~R.}\ \bibnamefont {White}} (\bibinfo {collaboration} {The Simons Collaboration on the Many-Electron Problem}),\ }\bibfield  {title} {\bibinfo {title} {Plaquette versus ordinary $d$-wave pairing in the ${t}^{\ensuremath{'}}$-hubbard model on a width-4 cylinder},\ }\href {https://doi.org/10.1103/PhysRevB.102.041106} {\bibfield  {journal} {\bibinfo  {journal} {Phys. Rev. B}\ }\textbf {\bibinfo {volume} {102}},\ \bibinfo {pages} {041106} (\bibinfo {year} {2020})}\BibitemShut {NoStop}%
\bibitem [{\citenamefont {Jiang}\ \emph {et~al.}(2020)\citenamefont {Jiang}, \citenamefont {Zaanen}, \citenamefont {Devereaux},\ and\ \citenamefont {Jiang}}]{PhysRevResearch.2.033073}%
  \BibitemOpen
  \bibfield  {author} {\bibinfo {author} {\bibfnamefont {Y.-F.}\ \bibnamefont {Jiang}}, \bibinfo {author} {\bibfnamefont {J.}~\bibnamefont {Zaanen}}, \bibinfo {author} {\bibfnamefont {T.~P.}\ \bibnamefont {Devereaux}},\ and\ \bibinfo {author} {\bibfnamefont {H.-C.}\ \bibnamefont {Jiang}},\ }\bibfield  {title} {\bibinfo {title} {Ground state phase diagram of the doped hubbard model on the four-leg cylinder},\ }\href {https://doi.org/10.1103/PhysRevResearch.2.033073} {\bibfield  {journal} {\bibinfo  {journal} {Phys. Rev. Res.}\ }\textbf {\bibinfo {volume} {2}},\ \bibinfo {pages} {033073} (\bibinfo {year} {2020})}\BibitemShut {NoStop}%
\bibitem [{\citenamefont {Danilov}\ \emph {et~al.}(2022)\citenamefont {Danilov}, \citenamefont {van Loon}, \citenamefont {Brener}, \citenamefont {Iskakov}, \citenamefont {Katsnelson},\ and\ \citenamefont {Lichtenstein}}]{danilov2022degenerate}%
  \BibitemOpen
  \bibfield  {author} {\bibinfo {author} {\bibfnamefont {M.}~\bibnamefont {Danilov}}, \bibinfo {author} {\bibfnamefont {E.~G.}\ \bibnamefont {van Loon}}, \bibinfo {author} {\bibfnamefont {S.}~\bibnamefont {Brener}}, \bibinfo {author} {\bibfnamefont {S.}~\bibnamefont {Iskakov}}, \bibinfo {author} {\bibfnamefont {M.~I.}\ \bibnamefont {Katsnelson}},\ and\ \bibinfo {author} {\bibfnamefont {A.~I.}\ \bibnamefont {Lichtenstein}},\ }\bibfield  {title} {\bibinfo {title} {Degenerate plaquette physics as key ingredient of high-temperature superconductivity in cuprates},\ }\href@noop {} {\bibfield  {journal} {\bibinfo  {journal} {npj Quantum Materials}\ }\textbf {\bibinfo {volume} {7}},\ \bibinfo {pages} {50} (\bibinfo {year} {2022})}\BibitemShut {NoStop}%
\bibitem [{\citenamefont {Lu}\ \emph {et~al.}(2024)\citenamefont {Lu}, \citenamefont {Chen}, \citenamefont {Zhu}, \citenamefont {Sheng},\ and\ \citenamefont {Gong}}]{PhysRevLett.132.066002}%
  \BibitemOpen
  \bibfield  {author} {\bibinfo {author} {\bibfnamefont {X.}~\bibnamefont {Lu}}, \bibinfo {author} {\bibfnamefont {F.}~\bibnamefont {Chen}}, \bibinfo {author} {\bibfnamefont {W.}~\bibnamefont {Zhu}}, \bibinfo {author} {\bibfnamefont {D.~N.}\ \bibnamefont {Sheng}},\ and\ \bibinfo {author} {\bibfnamefont {S.-S.}\ \bibnamefont {Gong}},\ }\bibfield  {title} {\bibinfo {title} {Emergent superconductivity and competing charge orders in hole-doped square-lattice {$t\text{\ensuremath{-}}J$} model},\ }\href {https://doi.org/10.1103/PhysRevLett.132.066002} {\bibfield  {journal} {\bibinfo  {journal} {Phys. Rev. Lett.}\ }\textbf {\bibinfo {volume} {132}},\ \bibinfo {pages} {066002} (\bibinfo {year} {2024})}\BibitemShut {NoStop}%
\bibitem [{\citenamefont {Jiang}\ \emph {et~al.}(2021)\citenamefont {Jiang}, \citenamefont {Scalapino},\ and\ \citenamefont {White}}]{doi:10.1073/pnas.2109978118}%
  \BibitemOpen
  \bibfield  {author} {\bibinfo {author} {\bibfnamefont {S.}~\bibnamefont {Jiang}}, \bibinfo {author} {\bibfnamefont {D.~J.}\ \bibnamefont {Scalapino}},\ and\ \bibinfo {author} {\bibfnamefont {S.~R.}\ \bibnamefont {White}},\ }\bibfield  {title} {\bibinfo {title} {Ground-state phase diagram of the {$t-t'-J$} model},\ }\href {https://doi.org/10.1073/pnas.2109978118} {\bibfield  {journal} {\bibinfo  {journal} {Proceedings of the National Academy of Sciences}\ }\textbf {\bibinfo {volume} {118}},\ \bibinfo {pages} {e2109978118} (\bibinfo {year} {2021})},\ \Eprint {https://arxiv.org/abs/https://www.pnas.org/doi/pdf/10.1073/pnas.2109978118} {https://www.pnas.org/doi/pdf/10.1073/pnas.2109978118} \BibitemShut {NoStop}%
\bibitem [{\citenamefont {Chen}\ \emph {et~al.}(2025)\citenamefont {Chen}, \citenamefont {Haldane},\ and\ \citenamefont {Sheng}}]{doi:10.1073/pnas.2420963122}%
  \BibitemOpen
  \bibfield  {author} {\bibinfo {author} {\bibfnamefont {F.}~\bibnamefont {Chen}}, \bibinfo {author} {\bibfnamefont {F.~D.~M.}\ \bibnamefont {Haldane}},\ and\ \bibinfo {author} {\bibfnamefont {D.~N.}\ \bibnamefont {Sheng}},\ }\bibfield  {title} {\bibinfo {title} {Global phase diagram of d-wave superconductivity in the square-lattice {$t-J$} model},\ }\href {https://doi.org/10.1073/pnas.2420963122} {\bibfield  {journal} {\bibinfo  {journal} {Proceedings of the National Academy of Sciences}\ }\textbf {\bibinfo {volume} {122}},\ \bibinfo {pages} {e2420963122} (\bibinfo {year} {2025})},\ \Eprint {https://arxiv.org/abs/https://www.pnas.org/doi/pdf/10.1073/pnas.2420963122} {https://www.pnas.org/doi/pdf/10.1073/pnas.2420963122} \BibitemShut {NoStop}%
\bibitem [{\citenamefont {Jiang}\ \emph {et~al.}(2012)\citenamefont {Jiang}, \citenamefont {Yao},\ and\ \citenamefont {Balents}}]{PhysRevB.86.024424}%
  \BibitemOpen
  \bibfield  {author} {\bibinfo {author} {\bibfnamefont {H.-C.}\ \bibnamefont {Jiang}}, \bibinfo {author} {\bibfnamefont {H.}~\bibnamefont {Yao}},\ and\ \bibinfo {author} {\bibfnamefont {L.}~\bibnamefont {Balents}},\ }\bibfield  {title} {\bibinfo {title} {Spin liquid ground state of the spin-$\frac{1}{2}$ square ${J}_{1}$-${J}_{2}$ heisenberg model},\ }\href {https://doi.org/10.1103/PhysRevB.86.024424} {\bibfield  {journal} {\bibinfo  {journal} {Phys. Rev. B}\ }\textbf {\bibinfo {volume} {86}},\ \bibinfo {pages} {024424} (\bibinfo {year} {2012})}\BibitemShut {NoStop}%
\bibitem [{\citenamefont {Nomura}\ and\ \citenamefont {Imada}(2021)}]{PhysRevX.11.031034}%
  \BibitemOpen
  \bibfield  {author} {\bibinfo {author} {\bibfnamefont {Y.}~\bibnamefont {Nomura}}\ and\ \bibinfo {author} {\bibfnamefont {M.}~\bibnamefont {Imada}},\ }\bibfield  {title} {\bibinfo {title} {Dirac-type nodal spin liquid revealed by refined quantum many-body solver using neural-network wave function, correlation ratio, and level spectroscopy},\ }\href {https://doi.org/10.1103/PhysRevX.11.031034} {\bibfield  {journal} {\bibinfo  {journal} {Phys. Rev. X}\ }\textbf {\bibinfo {volume} {11}},\ \bibinfo {pages} {031034} (\bibinfo {year} {2021})}\BibitemShut {NoStop}%
\bibitem [{\citenamefont {Qian}\ and\ \citenamefont {Qin}(2024)}]{PhysRevB.109.L161103}%
  \BibitemOpen
  \bibfield  {author} {\bibinfo {author} {\bibfnamefont {X.}~\bibnamefont {Qian}}\ and\ \bibinfo {author} {\bibfnamefont {M.}~\bibnamefont {Qin}},\ }\bibfield  {title} {\bibinfo {title} {Absence of spin liquid phase in the ${J}_{1}\ensuremath{-}{J}_{2}$ heisenberg model on the square lattice},\ }\href {https://doi.org/10.1103/PhysRevB.109.L161103} {\bibfield  {journal} {\bibinfo  {journal} {Phys. Rev. B}\ }\textbf {\bibinfo {volume} {109}},\ \bibinfo {pages} {L161103} (\bibinfo {year} {2024})}\BibitemShut {NoStop}%
\bibitem [{\citenamefont {R\"uckriegel}\ \emph {et~al.}(2024)\citenamefont {R\"uckriegel}, \citenamefont {Tarasevych},\ and\ \citenamefont {Kopietz}}]{PhysRevB.109.184410}%
  \BibitemOpen
  \bibfield  {author} {\bibinfo {author} {\bibfnamefont {A.}~\bibnamefont {R\"uckriegel}}, \bibinfo {author} {\bibfnamefont {D.}~\bibnamefont {Tarasevych}},\ and\ \bibinfo {author} {\bibfnamefont {P.}~\bibnamefont {Kopietz}},\ }\bibfield  {title} {\bibinfo {title} {Phase diagram of the ${J}_{1}\text{\ensuremath{-}}{J}_{2}$ quantum heisenberg model for arbitrary spin},\ }\href {https://doi.org/10.1103/PhysRevB.109.184410} {\bibfield  {journal} {\bibinfo  {journal} {Phys. Rev. B}\ }\textbf {\bibinfo {volume} {109}},\ \bibinfo {pages} {184410} (\bibinfo {year} {2024})}\BibitemShut {NoStop}%
\bibitem [{\citenamefont {Kariyado}(2023)}]{PhysRevB.107.085127}%
  \BibitemOpen
  \bibfield  {author} {\bibinfo {author} {\bibfnamefont {T.}~\bibnamefont {Kariyado}},\ }\bibfield  {title} {\bibinfo {title} {Twisted bilayer ${\text{bc}}_{3}$: Valley interlocked anisotropic flat bands},\ }\href {https://doi.org/10.1103/PhysRevB.107.085127} {\bibfield  {journal} {\bibinfo  {journal} {Phys. Rev. B}\ }\textbf {\bibinfo {volume} {107}},\ \bibinfo {pages} {085127} (\bibinfo {year} {2023})}\BibitemShut {NoStop}%
\bibitem [{\citenamefont {Volkov}\ \emph {et~al.}(2023{\natexlab{a}})\citenamefont {Volkov}, \citenamefont {Wilson}, \citenamefont {Lucht},\ and\ \citenamefont {Pixley}}]{PhysRevB.107.174506}%
  \BibitemOpen
  \bibfield  {author} {\bibinfo {author} {\bibfnamefont {P.~A.}\ \bibnamefont {Volkov}}, \bibinfo {author} {\bibfnamefont {J.~H.}\ \bibnamefont {Wilson}}, \bibinfo {author} {\bibfnamefont {K.~P.}\ \bibnamefont {Lucht}},\ and\ \bibinfo {author} {\bibfnamefont {J.~H.}\ \bibnamefont {Pixley}},\ }\bibfield  {title} {\bibinfo {title} {Magic angles and correlations in twisted nodal superconductors},\ }\href {https://doi.org/10.1103/PhysRevB.107.174506} {\bibfield  {journal} {\bibinfo  {journal} {Phys. Rev. B}\ }\textbf {\bibinfo {volume} {107}},\ \bibinfo {pages} {174506} (\bibinfo {year} {2023}{\natexlab{a}})}\BibitemShut {NoStop}%
\bibitem [{\citenamefont {Soeda}\ \emph {et~al.}(2022)\citenamefont {Soeda}, \citenamefont {Asaga},\ and\ \citenamefont {Fukui}}]{PhysRevB.105.165422}%
  \BibitemOpen
  \bibfield  {author} {\bibinfo {author} {\bibfnamefont {Y.}~\bibnamefont {Soeda}}, \bibinfo {author} {\bibfnamefont {K.}~\bibnamefont {Asaga}},\ and\ \bibinfo {author} {\bibfnamefont {T.}~\bibnamefont {Fukui}},\ }\bibfield  {title} {\bibinfo {title} {Moir\'e landau levels of a ${C}_{4}$-symmetric twisted bilayer system in the absence of a magnetic field},\ }\href {https://doi.org/10.1103/PhysRevB.105.165422} {\bibfield  {journal} {\bibinfo  {journal} {Phys. Rev. B}\ }\textbf {\bibinfo {volume} {105}},\ \bibinfo {pages} {165422} (\bibinfo {year} {2022})}\BibitemShut {NoStop}%
\bibitem [{\citenamefont {Luo}\ \emph {et~al.}(2021)\citenamefont {Luo}, \citenamefont {Xu},\ and\ \citenamefont {Jian}}]{PhysRevB.104.035136}%
  \BibitemOpen
  \bibfield  {author} {\bibinfo {author} {\bibfnamefont {Z.-X.}\ \bibnamefont {Luo}}, \bibinfo {author} {\bibfnamefont {C.}~\bibnamefont {Xu}},\ and\ \bibinfo {author} {\bibfnamefont {C.-M.}\ \bibnamefont {Jian}},\ }\bibfield  {title} {\bibinfo {title} {Magic continuum in a twisted bilayer square lattice with staggered flux},\ }\href {https://doi.org/10.1103/PhysRevB.104.035136} {\bibfield  {journal} {\bibinfo  {journal} {Phys. Rev. B}\ }\textbf {\bibinfo {volume} {104}},\ \bibinfo {pages} {035136} (\bibinfo {year} {2021})}\BibitemShut {NoStop}%
\bibitem [{\citenamefont {Li}\ \emph {et~al.}(2022)\citenamefont {Li}, \citenamefont {He},\ and\ \citenamefont {Yao}}]{PhysRevResearch.4.043151}%
  \BibitemOpen
  \bibfield  {author} {\bibinfo {author} {\bibfnamefont {M.-R.}\ \bibnamefont {Li}}, \bibinfo {author} {\bibfnamefont {A.-L.}\ \bibnamefont {He}},\ and\ \bibinfo {author} {\bibfnamefont {H.}~\bibnamefont {Yao}},\ }\bibfield  {title} {\bibinfo {title} {Magic-angle twisted bilayer systems with quadratic band touching: Exactly flat bands with high chern number},\ }\href {https://doi.org/10.1103/PhysRevResearch.4.043151} {\bibfield  {journal} {\bibinfo  {journal} {Phys. Rev. Res.}\ }\textbf {\bibinfo {volume} {4}},\ \bibinfo {pages} {043151} (\bibinfo {year} {2022})}\BibitemShut {NoStop}%
\bibitem [{\citenamefont {Can}\ \emph {et~al.}(2021)\citenamefont {Can}, \citenamefont {Tummuru}, \citenamefont {Day}, \citenamefont {Elfimov}, \citenamefont {Damascelli},\ and\ \citenamefont {Franz}}]{can2021high}%
  \BibitemOpen
  \bibfield  {author} {\bibinfo {author} {\bibfnamefont {O.}~\bibnamefont {Can}}, \bibinfo {author} {\bibfnamefont {T.}~\bibnamefont {Tummuru}}, \bibinfo {author} {\bibfnamefont {R.~P.}\ \bibnamefont {Day}}, \bibinfo {author} {\bibfnamefont {I.}~\bibnamefont {Elfimov}}, \bibinfo {author} {\bibfnamefont {A.}~\bibnamefont {Damascelli}},\ and\ \bibinfo {author} {\bibfnamefont {M.}~\bibnamefont {Franz}},\ }\bibfield  {title} {\bibinfo {title} {High-temperature topological superconductivity in twisted double-layer copper oxides},\ }\href@noop {} {\bibfield  {journal} {\bibinfo  {journal} {Nature Physics}\ }\textbf {\bibinfo {volume} {17}},\ \bibinfo {pages} {519} (\bibinfo {year} {2021})}\BibitemShut {NoStop}%
\bibitem [{\citenamefont {Zhao}\ \emph {et~al.}(2023{\natexlab{b}})\citenamefont {Zhao}, \citenamefont {Cui}, \citenamefont {Volkov}, \citenamefont {Yoo}, \citenamefont {Lee}, \citenamefont {Gardener}, \citenamefont {Akey}, \citenamefont {Engelke}, \citenamefont {Ronen}, \citenamefont {Zhong}, \citenamefont {Gu}, \citenamefont {Plugge}, \citenamefont {Tummuru}, \citenamefont {Kim}, \citenamefont {Franz}, \citenamefont {Pixley}, \citenamefont {Poccia},\ and\ \citenamefont {Kim}}]{doi:10.1126/science.abl8371}%
  \BibitemOpen
  \bibfield  {author} {\bibinfo {author} {\bibfnamefont {S.~Y.~F.}\ \bibnamefont {Zhao}}, \bibinfo {author} {\bibfnamefont {X.}~\bibnamefont {Cui}}, \bibinfo {author} {\bibfnamefont {P.~A.}\ \bibnamefont {Volkov}}, \bibinfo {author} {\bibfnamefont {H.}~\bibnamefont {Yoo}}, \bibinfo {author} {\bibfnamefont {S.}~\bibnamefont {Lee}}, \bibinfo {author} {\bibfnamefont {J.~A.}\ \bibnamefont {Gardener}}, \bibinfo {author} {\bibfnamefont {A.~J.}\ \bibnamefont {Akey}}, \bibinfo {author} {\bibfnamefont {R.}~\bibnamefont {Engelke}}, \bibinfo {author} {\bibfnamefont {Y.}~\bibnamefont {Ronen}}, \bibinfo {author} {\bibfnamefont {R.}~\bibnamefont {Zhong}}, \bibinfo {author} {\bibfnamefont {G.}~\bibnamefont {Gu}}, \bibinfo {author} {\bibfnamefont {S.}~\bibnamefont {Plugge}}, \bibinfo {author} {\bibfnamefont {T.}~\bibnamefont {Tummuru}}, \bibinfo {author} {\bibfnamefont {M.}~\bibnamefont {Kim}}, \bibinfo {author} {\bibfnamefont {M.}~\bibnamefont {Franz}}, \bibinfo {author} {\bibfnamefont {J.~H.}\ \bibnamefont {Pixley}},
  \bibinfo {author} {\bibfnamefont {N.}~\bibnamefont {Poccia}},\ and\ \bibinfo {author} {\bibfnamefont {P.}~\bibnamefont {Kim}},\ }\bibfield  {title} {\bibinfo {title} {Time-reversal symmetry breaking superconductivity between twisted cuprate superconductors},\ }\href {https://doi.org/10.1126/science.abl8371} {\bibfield  {journal} {\bibinfo  {journal} {Science}\ }\textbf {\bibinfo {volume} {382}},\ \bibinfo {pages} {1422} (\bibinfo {year} {2023}{\natexlab{b}})},\ \Eprint {https://arxiv.org/abs/https://www.science.org/doi/pdf/10.1126/science.abl8371} {https://www.science.org/doi/pdf/10.1126/science.abl8371} \BibitemShut {NoStop}%
\bibitem [{\citenamefont {Song}\ \emph {et~al.}(2022)\citenamefont {Song}, \citenamefont {Zhang},\ and\ \citenamefont {Vishwanath}}]{PhysRevB.105.L201102}%
  \BibitemOpen
  \bibfield  {author} {\bibinfo {author} {\bibfnamefont {X.-Y.}\ \bibnamefont {Song}}, \bibinfo {author} {\bibfnamefont {Y.-H.}\ \bibnamefont {Zhang}},\ and\ \bibinfo {author} {\bibfnamefont {A.}~\bibnamefont {Vishwanath}},\ }\bibfield  {title} {\bibinfo {title} {Doping a moir\'e mott insulator: A {$t\ensuremath{-}J$} model study of twisted cuprates},\ }\href {https://doi.org/10.1103/PhysRevB.105.L201102} {\bibfield  {journal} {\bibinfo  {journal} {Phys. Rev. B}\ }\textbf {\bibinfo {volume} {105}},\ \bibinfo {pages} {L201102} (\bibinfo {year} {2022})}\BibitemShut {NoStop}%
\bibitem [{\citenamefont {Haenel}\ \emph {et~al.}(2022)\citenamefont {Haenel}, \citenamefont {Tummuru},\ and\ \citenamefont {Franz}}]{PhysRevB.106.104505}%
  \BibitemOpen
  \bibfield  {author} {\bibinfo {author} {\bibfnamefont {R.}~\bibnamefont {Haenel}}, \bibinfo {author} {\bibfnamefont {T.}~\bibnamefont {Tummuru}},\ and\ \bibinfo {author} {\bibfnamefont {M.}~\bibnamefont {Franz}},\ }\bibfield  {title} {\bibinfo {title} {Incoherent tunneling and topological superconductivity in twisted cuprate bilayers},\ }\href {https://doi.org/10.1103/PhysRevB.106.104505} {\bibfield  {journal} {\bibinfo  {journal} {Phys. Rev. B}\ }\textbf {\bibinfo {volume} {106}},\ \bibinfo {pages} {104505} (\bibinfo {year} {2022})}\BibitemShut {NoStop}%
\bibitem [{\citenamefont {Volkov}\ \emph {et~al.}(2023{\natexlab{b}})\citenamefont {Volkov}, \citenamefont {Wilson}, \citenamefont {Lucht},\ and\ \citenamefont {Pixley}}]{PhysRevLett.130.186001}%
  \BibitemOpen
  \bibfield  {author} {\bibinfo {author} {\bibfnamefont {P.~A.}\ \bibnamefont {Volkov}}, \bibinfo {author} {\bibfnamefont {J.~H.}\ \bibnamefont {Wilson}}, \bibinfo {author} {\bibfnamefont {K.~P.}\ \bibnamefont {Lucht}},\ and\ \bibinfo {author} {\bibfnamefont {J.~H.}\ \bibnamefont {Pixley}},\ }\bibfield  {title} {\bibinfo {title} {Current- and field-induced topology in twisted nodal superconductors},\ }\href {https://doi.org/10.1103/PhysRevLett.130.186001} {\bibfield  {journal} {\bibinfo  {journal} {Phys. Rev. Lett.}\ }\textbf {\bibinfo {volume} {130}},\ \bibinfo {pages} {186001} (\bibinfo {year} {2023}{\natexlab{b}})}\BibitemShut {NoStop}%
\bibitem [{\citenamefont {Eugenio}\ and\ \citenamefont {Vafek}(2023)}]{10.21468/SciPostPhys.15.3.081}%
  \BibitemOpen
  \bibfield  {author} {\bibinfo {author} {\bibfnamefont {P.~M.}\ \bibnamefont {Eugenio}}\ and\ \bibinfo {author} {\bibfnamefont {O.}~\bibnamefont {Vafek}},\ }\bibfield  {title} {\bibinfo {title} {{Twisted-bilayer FeSe and the Fe-based superlattices}},\ }\href {https://doi.org/10.21468/SciPostPhys.15.3.081} {\bibfield  {journal} {\bibinfo  {journal} {SciPost Phys.}\ }\textbf {\bibinfo {volume} {15}},\ \bibinfo {pages} {081} (\bibinfo {year} {2023})}\BibitemShut {NoStop}%
\bibitem [{\citenamefont {Akashi}\ \emph {et~al.}(2017)\citenamefont {Akashi}, \citenamefont {Iida}, \citenamefont {Yamamoto},\ and\ \citenamefont {Yoshizawa}}]{PhysRevB.95.245401}%
  \BibitemOpen
  \bibfield  {author} {\bibinfo {author} {\bibfnamefont {R.}~\bibnamefont {Akashi}}, \bibinfo {author} {\bibfnamefont {Y.}~\bibnamefont {Iida}}, \bibinfo {author} {\bibfnamefont {K.}~\bibnamefont {Yamamoto}},\ and\ \bibinfo {author} {\bibfnamefont {K.}~\bibnamefont {Yoshizawa}},\ }\bibfield  {title} {\bibinfo {title} {Interference of the {Bloch} phase in layered materials with stacking shifts},\ }\href {https://doi.org/10.1103/PhysRevB.95.245401} {\bibfield  {journal} {\bibinfo  {journal} {Phys. Rev. B}\ }\textbf {\bibinfo {volume} {95}},\ \bibinfo {pages} {245401} (\bibinfo {year} {2017})}\BibitemShut {NoStop}%
\bibitem [{\citenamefont {Marzari}\ and\ \citenamefont {Vanderbilt}(1997)}]{PhysRevB.56.12847}%
  \BibitemOpen
  \bibfield  {author} {\bibinfo {author} {\bibfnamefont {N.}~\bibnamefont {Marzari}}\ and\ \bibinfo {author} {\bibfnamefont {D.}~\bibnamefont {Vanderbilt}},\ }\bibfield  {title} {\bibinfo {title} {Maximally localized generalized {Wannier} functions for composite energy bands},\ }\href {https://doi.org/10.1103/PhysRevB.56.12847} {\bibfield  {journal} {\bibinfo  {journal} {Phys. Rev. B}\ }\textbf {\bibinfo {volume} {56}},\ \bibinfo {pages} {12847} (\bibinfo {year} {1997})}\BibitemShut {NoStop}%
\bibitem [{\citenamefont {Mazurenko}\ \emph {et~al.}(2007)\citenamefont {Mazurenko}, \citenamefont {Skornyakov}, \citenamefont {Kozhevnikov}, \citenamefont {Mila},\ and\ \citenamefont {Anisimov}}]{PhysRevB.75.224408}%
  \BibitemOpen
  \bibfield  {author} {\bibinfo {author} {\bibfnamefont {V.~V.}\ \bibnamefont {Mazurenko}}, \bibinfo {author} {\bibfnamefont {S.~L.}\ \bibnamefont {Skornyakov}}, \bibinfo {author} {\bibfnamefont {A.~V.}\ \bibnamefont {Kozhevnikov}}, \bibinfo {author} {\bibfnamefont {F.}~\bibnamefont {Mila}},\ and\ \bibinfo {author} {\bibfnamefont {V.~I.}\ \bibnamefont {Anisimov}},\ }\bibfield  {title} {\bibinfo {title} {Wannier functions and exchange integrals: The example of $\mathrm{Li}{\mathrm{cu}}_{2}{\mathrm{o}}_{2}$},\ }\href {https://doi.org/10.1103/PhysRevB.75.224408} {\bibfield  {journal} {\bibinfo  {journal} {Phys. Rev. B}\ }\textbf {\bibinfo {volume} {75}},\ \bibinfo {pages} {224408} (\bibinfo {year} {2007})}\BibitemShut {NoStop}%
\bibitem [{\citenamefont {Cea}\ \emph {et~al.}(2019)\citenamefont {Cea}, \citenamefont {Walet},\ and\ \citenamefont {Guinea}}]{PhysRevB.100.205113}%
  \BibitemOpen
  \bibfield  {author} {\bibinfo {author} {\bibfnamefont {T.}~\bibnamefont {Cea}}, \bibinfo {author} {\bibfnamefont {N.~R.}\ \bibnamefont {Walet}},\ and\ \bibinfo {author} {\bibfnamefont {F.}~\bibnamefont {Guinea}},\ }\bibfield  {title} {\bibinfo {title} {Electronic band structure and pinning of {Fermi} energy to {Van Hove} singularities in twisted bilayer graphene: A self-consistent approach},\ }\href {https://doi.org/10.1103/PhysRevB.100.205113} {\bibfield  {journal} {\bibinfo  {journal} {Phys. Rev. B}\ }\textbf {\bibinfo {volume} {100}},\ \bibinfo {pages} {205113} (\bibinfo {year} {2019})}\BibitemShut {NoStop}%
\bibitem [{\citenamefont {Kang}\ and\ \citenamefont {Vafek}(2020)}]{PhysRevB.102.035161}%
  \BibitemOpen
  \bibfield  {author} {\bibinfo {author} {\bibfnamefont {J.}~\bibnamefont {Kang}}\ and\ \bibinfo {author} {\bibfnamefont {O.}~\bibnamefont {Vafek}},\ }\bibfield  {title} {\bibinfo {title} {Non-{Abelian} {Dirac} node braiding and near-degeneracy of correlated phases at odd integer filling in magic-angle twisted bilayer graphene},\ }\href {https://doi.org/10.1103/PhysRevB.102.035161} {\bibfield  {journal} {\bibinfo  {journal} {Phys. Rev. B}\ }\textbf {\bibinfo {volume} {102}},\ \bibinfo {pages} {035161} (\bibinfo {year} {2020})}\BibitemShut {NoStop}%
\bibitem [{\citenamefont {Cea}\ and\ \citenamefont {Guinea}(2020)}]{PhysRevB.102.045107}%
  \BibitemOpen
  \bibfield  {author} {\bibinfo {author} {\bibfnamefont {T.}~\bibnamefont {Cea}}\ and\ \bibinfo {author} {\bibfnamefont {F.}~\bibnamefont {Guinea}},\ }\bibfield  {title} {\bibinfo {title} {Band structure and insulating states driven by {Coulomb} interaction in twisted bilayer graphene},\ }\href {https://doi.org/10.1103/PhysRevB.102.045107} {\bibfield  {journal} {\bibinfo  {journal} {Phys. Rev. B}\ }\textbf {\bibinfo {volume} {102}},\ \bibinfo {pages} {045107} (\bibinfo {year} {2020})}\BibitemShut {NoStop}%
\bibitem [{\citenamefont {Li}\ \emph {et~al.}(2010)\citenamefont {Li}, \citenamefont {Li}, \citenamefont {Liu}, \citenamefont {Guo}, \citenamefont {Li},\ and\ \citenamefont {Zhu}}]{B922733D}%
  \BibitemOpen
  \bibfield  {author} {\bibinfo {author} {\bibfnamefont {G.}~\bibnamefont {Li}}, \bibinfo {author} {\bibfnamefont {Y.}~\bibnamefont {Li}}, \bibinfo {author} {\bibfnamefont {H.}~\bibnamefont {Liu}}, \bibinfo {author} {\bibfnamefont {Y.}~\bibnamefont {Guo}}, \bibinfo {author} {\bibfnamefont {Y.}~\bibnamefont {Li}},\ and\ \bibinfo {author} {\bibfnamefont {D.}~\bibnamefont {Zhu}},\ }\bibfield  {title} {\bibinfo {title} {Architecture of graphdiyne nanoscale films},\ }\href {https://doi.org/10.1039/B922733D} {\bibfield  {journal} {\bibinfo  {journal} {Chem. Commun.}\ }\textbf {\bibinfo {volume} {46}},\ \bibinfo {pages} {3256} (\bibinfo {year} {2010})}\BibitemShut {NoStop}%
\bibitem [{\citenamefont {Giannozzi}\ \emph {et~al.}(2009)\citenamefont {Giannozzi}, \citenamefont {Baroni}, \citenamefont {Bonini}, \citenamefont {Calandra}, \citenamefont {Car}, \citenamefont {Cavazzoni}, \citenamefont {Ceresoli}, \citenamefont {Chiarotti}, \citenamefont {Cococcioni}, \citenamefont {Dabo}, \citenamefont {Corso}, \citenamefont {de~Gironcoli}, \citenamefont {Fabris}, \citenamefont {Fratesi}, \citenamefont {Gebauer}, \citenamefont {Gerstmann}, \citenamefont {Gougoussis}, \citenamefont {Kokalj}, \citenamefont {Lazzeri}, \citenamefont {Martin-Samos}, \citenamefont {Marzari}, \citenamefont {Mauri}, \citenamefont {Mazzarello}, \citenamefont {Paolini}, \citenamefont {Pasquarello}, \citenamefont {Paulatto}, \citenamefont {Sbraccia}, \citenamefont {Scandolo}, \citenamefont {Sclauzero}, \citenamefont {Seitsonen}, \citenamefont {Smogunov}, \citenamefont {Umari},\ and\ \citenamefont {Wentzcovitch}}]{Giannozzi_2009}%
  \BibitemOpen
  \bibfield  {author} {\bibinfo {author} {\bibfnamefont {P.}~\bibnamefont {Giannozzi}}, \bibinfo {author} {\bibfnamefont {S.}~\bibnamefont {Baroni}}, \bibinfo {author} {\bibfnamefont {N.}~\bibnamefont {Bonini}}, \bibinfo {author} {\bibfnamefont {M.}~\bibnamefont {Calandra}}, \bibinfo {author} {\bibfnamefont {R.}~\bibnamefont {Car}}, \bibinfo {author} {\bibfnamefont {C.}~\bibnamefont {Cavazzoni}}, \bibinfo {author} {\bibfnamefont {D.}~\bibnamefont {Ceresoli}}, \bibinfo {author} {\bibfnamefont {G.~L.}\ \bibnamefont {Chiarotti}}, \bibinfo {author} {\bibfnamefont {M.}~\bibnamefont {Cococcioni}}, \bibinfo {author} {\bibfnamefont {I.}~\bibnamefont {Dabo}}, \bibinfo {author} {\bibfnamefont {A.~D.}\ \bibnamefont {Corso}}, \bibinfo {author} {\bibfnamefont {S.}~\bibnamefont {de~Gironcoli}}, \bibinfo {author} {\bibfnamefont {S.}~\bibnamefont {Fabris}}, \bibinfo {author} {\bibfnamefont {G.}~\bibnamefont {Fratesi}}, \bibinfo {author} {\bibfnamefont {R.}~\bibnamefont {Gebauer}}, \bibinfo {author} {\bibfnamefont
  {U.}~\bibnamefont {Gerstmann}}, \bibinfo {author} {\bibfnamefont {C.}~\bibnamefont {Gougoussis}}, \bibinfo {author} {\bibfnamefont {A.}~\bibnamefont {Kokalj}}, \bibinfo {author} {\bibfnamefont {M.}~\bibnamefont {Lazzeri}}, \bibinfo {author} {\bibfnamefont {L.}~\bibnamefont {Martin-Samos}}, \bibinfo {author} {\bibfnamefont {N.}~\bibnamefont {Marzari}}, \bibinfo {author} {\bibfnamefont {F.}~\bibnamefont {Mauri}}, \bibinfo {author} {\bibfnamefont {R.}~\bibnamefont {Mazzarello}}, \bibinfo {author} {\bibfnamefont {S.}~\bibnamefont {Paolini}}, \bibinfo {author} {\bibfnamefont {A.}~\bibnamefont {Pasquarello}}, \bibinfo {author} {\bibfnamefont {L.}~\bibnamefont {Paulatto}}, \bibinfo {author} {\bibfnamefont {C.}~\bibnamefont {Sbraccia}}, \bibinfo {author} {\bibfnamefont {S.}~\bibnamefont {Scandolo}}, \bibinfo {author} {\bibfnamefont {G.}~\bibnamefont {Sclauzero}}, \bibinfo {author} {\bibfnamefont {A.~P.}\ \bibnamefont {Seitsonen}}, \bibinfo {author} {\bibfnamefont {A.}~\bibnamefont {Smogunov}}, \bibinfo {author}
  {\bibfnamefont {P.}~\bibnamefont {Umari}},\ and\ \bibinfo {author} {\bibfnamefont {R.~M.}\ \bibnamefont {Wentzcovitch}},\ }\bibfield  {title} {\bibinfo {title} {{QUANTUM} {ESPRESSO}: a modular and open-source software project for quantum simulations of materials},\ }\href {https://doi.org/10.1088/0953-8984/21/39/395502} {\bibfield  {journal} {\bibinfo  {journal} {J. Phys. Condens. Matter}\ }\textbf {\bibinfo {volume} {21}},\ \bibinfo {pages} {395502} (\bibinfo {year} {2009})}\BibitemShut {NoStop}%
\bibitem [{\citenamefont {Giannozzi}\ \emph {et~al.}(2017)\citenamefont {Giannozzi}, \citenamefont {Andreussi}, \citenamefont {Brumme}, \citenamefont {Bunau}, \citenamefont {Nardelli}, \citenamefont {Calandra}, \citenamefont {Car}, \citenamefont {Cavazzoni}, \citenamefont {Ceresoli}, \citenamefont {Cococcioni}, \citenamefont {Colonna}, \citenamefont {Carnimeo}, \citenamefont {Corso}, \citenamefont {de~Gironcoli}, \citenamefont {Delugas}, \citenamefont {DiStasio}, \citenamefont {Ferretti}, \citenamefont {Floris}, \citenamefont {Fratesi}, \citenamefont {Fugallo}, \citenamefont {Gebauer}, \citenamefont {Gerstmann}, \citenamefont {Giustino}, \citenamefont {Gorni}, \citenamefont {Jia}, \citenamefont {Kawamura}, \citenamefont {Ko}, \citenamefont {Kokalj}, \citenamefont {Kü{\c{c}}ükbenli}, \citenamefont {Lazzeri}, \citenamefont {Marsili}, \citenamefont {Marzari}, \citenamefont {Mauri}, \citenamefont {Nguyen}, \citenamefont {Nguyen}, \citenamefont {de-la Roza}, \citenamefont {Paulatto}, \citenamefont {Ponc{\'{e}}},
  \citenamefont {Rocca}, \citenamefont {Sabatini}, \citenamefont {Santra}, \citenamefont {Schlipf}, \citenamefont {Seitsonen}, \citenamefont {Smogunov}, \citenamefont {Timrov}, \citenamefont {Thonhauser}, \citenamefont {Umari}, \citenamefont {Vast}, \citenamefont {Wu},\ and\ \citenamefont {Baroni}}]{Giannozzi_2017}%
  \BibitemOpen
  \bibfield  {author} {\bibinfo {author} {\bibfnamefont {P.}~\bibnamefont {Giannozzi}}, \bibinfo {author} {\bibfnamefont {O.}~\bibnamefont {Andreussi}}, \bibinfo {author} {\bibfnamefont {T.}~\bibnamefont {Brumme}}, \bibinfo {author} {\bibfnamefont {O.}~\bibnamefont {Bunau}}, \bibinfo {author} {\bibfnamefont {M.~B.}\ \bibnamefont {Nardelli}}, \bibinfo {author} {\bibfnamefont {M.}~\bibnamefont {Calandra}}, \bibinfo {author} {\bibfnamefont {R.}~\bibnamefont {Car}}, \bibinfo {author} {\bibfnamefont {C.}~\bibnamefont {Cavazzoni}}, \bibinfo {author} {\bibfnamefont {D.}~\bibnamefont {Ceresoli}}, \bibinfo {author} {\bibfnamefont {M.}~\bibnamefont {Cococcioni}}, \bibinfo {author} {\bibfnamefont {N.}~\bibnamefont {Colonna}}, \bibinfo {author} {\bibfnamefont {I.}~\bibnamefont {Carnimeo}}, \bibinfo {author} {\bibfnamefont {A.~D.}\ \bibnamefont {Corso}}, \bibinfo {author} {\bibfnamefont {S.}~\bibnamefont {de~Gironcoli}}, \bibinfo {author} {\bibfnamefont {P.}~\bibnamefont {Delugas}}, \bibinfo {author} {\bibfnamefont
  {R.~A.}\ \bibnamefont {DiStasio}}, \bibinfo {author} {\bibfnamefont {A.}~\bibnamefont {Ferretti}}, \bibinfo {author} {\bibfnamefont {A.}~\bibnamefont {Floris}}, \bibinfo {author} {\bibfnamefont {G.}~\bibnamefont {Fratesi}}, \bibinfo {author} {\bibfnamefont {G.}~\bibnamefont {Fugallo}}, \bibinfo {author} {\bibfnamefont {R.}~\bibnamefont {Gebauer}}, \bibinfo {author} {\bibfnamefont {U.}~\bibnamefont {Gerstmann}}, \bibinfo {author} {\bibfnamefont {F.}~\bibnamefont {Giustino}}, \bibinfo {author} {\bibfnamefont {T.}~\bibnamefont {Gorni}}, \bibinfo {author} {\bibfnamefont {J.}~\bibnamefont {Jia}}, \bibinfo {author} {\bibfnamefont {M.}~\bibnamefont {Kawamura}}, \bibinfo {author} {\bibfnamefont {H.-Y.}\ \bibnamefont {Ko}}, \bibinfo {author} {\bibfnamefont {A.}~\bibnamefont {Kokalj}}, \bibinfo {author} {\bibfnamefont {E.}~\bibnamefont {Kü{\c{c}}ükbenli}}, \bibinfo {author} {\bibfnamefont {M.}~\bibnamefont {Lazzeri}}, \bibinfo {author} {\bibfnamefont {M.}~\bibnamefont {Marsili}}, \bibinfo {author} {\bibfnamefont
  {N.}~\bibnamefont {Marzari}}, \bibinfo {author} {\bibfnamefont {F.}~\bibnamefont {Mauri}}, \bibinfo {author} {\bibfnamefont {N.~L.}\ \bibnamefont {Nguyen}}, \bibinfo {author} {\bibfnamefont {H.-V.}\ \bibnamefont {Nguyen}}, \bibinfo {author} {\bibfnamefont {A.~O.}\ \bibnamefont {de-la Roza}}, \bibinfo {author} {\bibfnamefont {L.}~\bibnamefont {Paulatto}}, \bibinfo {author} {\bibfnamefont {S.}~\bibnamefont {Ponc{\'{e}}}}, \bibinfo {author} {\bibfnamefont {D.}~\bibnamefont {Rocca}}, \bibinfo {author} {\bibfnamefont {R.}~\bibnamefont {Sabatini}}, \bibinfo {author} {\bibfnamefont {B.}~\bibnamefont {Santra}}, \bibinfo {author} {\bibfnamefont {M.}~\bibnamefont {Schlipf}}, \bibinfo {author} {\bibfnamefont {A.~P.}\ \bibnamefont {Seitsonen}}, \bibinfo {author} {\bibfnamefont {A.}~\bibnamefont {Smogunov}}, \bibinfo {author} {\bibfnamefont {I.}~\bibnamefont {Timrov}}, \bibinfo {author} {\bibfnamefont {T.}~\bibnamefont {Thonhauser}}, \bibinfo {author} {\bibfnamefont {P.}~\bibnamefont {Umari}}, \bibinfo {author}
  {\bibfnamefont {N.}~\bibnamefont {Vast}}, \bibinfo {author} {\bibfnamefont {X.}~\bibnamefont {Wu}},\ and\ \bibinfo {author} {\bibfnamefont {S.}~\bibnamefont {Baroni}},\ }\bibfield  {title} {\bibinfo {title} {Advanced capabilities for materials modelling with {Quantum} {ESPRESSO}},\ }\href {https://doi.org/10.1088/1361-648x/aa8f79} {\bibfield  {journal} {\bibinfo  {journal} {J. Phys. Condens. Matter}\ }\textbf {\bibinfo {volume} {29}},\ \bibinfo {pages} {465901} (\bibinfo {year} {2017})}\BibitemShut {NoStop}%
\bibitem [{\citenamefont {Hamada}(2014)}]{PhysRevB.89.121103}%
  \BibitemOpen
  \bibfield  {author} {\bibinfo {author} {\bibfnamefont {I.}~\bibnamefont {Hamada}},\ }\bibfield  {title} {\bibinfo {title} {van der {Waals} density functional made accurate},\ }\href {https://doi.org/10.1103/PhysRevB.89.121103} {\bibfield  {journal} {\bibinfo  {journal} {Phys. Rev. B}\ }\textbf {\bibinfo {volume} {89}},\ \bibinfo {pages} {121103(R)} (\bibinfo {year} {2014})}\BibitemShut {NoStop}%
\bibitem [{\citenamefont {{Dal Corso}}(2014)}]{DALCORSO2014337}%
  \BibitemOpen
  \bibfield  {author} {\bibinfo {author} {\bibfnamefont {A.}~\bibnamefont {{Dal Corso}}},\ }\bibfield  {title} {\bibinfo {title} {Pseudopotentials periodic table: From {H} to {Pu}},\ }\href {https://doi.org/https://doi.org/10.1016/j.commatsci.2014.07.043} {\bibfield  {journal} {\bibinfo  {journal} {Comput. Mater. Sci.}\ }\textbf {\bibinfo {volume} {95}},\ \bibinfo {pages} {337} (\bibinfo {year} {2014})}\BibitemShut {NoStop}%
\bibitem [{psl()}]{pslibrary}%
  \BibitemOpen
  \href@noop {} {}\bibinfo {note} {{https://dalcorso.github.io/pslibrary/}}\BibitemShut {NoStop}%
\end{thebibliography}%
\end{document}